\documentclass[twocolumn,showpacs,preprintnumbers,amsmath,amssymb]{revtex4}
\usepackage{amsmath,amssymb,graphics,epsfig,subfigure}
\usepackage{color}

\begin{document}
\newcommand {\nn} {\nonumber}
\renewcommand{\baselinestretch}{1.3}

\title{Topology of black hole thermodynamics: A brief review}

\author{Shao-Wen Wei$^{1,2}$ \footnote{weishw@lzu.edu.cn},
        Yu-Xiao Liu$^{1,2}$\footnote{liuyx@lzu.edu.cn}}

\affiliation{$^{1}$Lanzhou Center for Theoretical Physics, Key Laboratory of Theoretical Physics of Gansu Province, and Key Laboratory of Quantum Theory and Applications of MoE, Lanzhou University, Lanzhou, Gansu 730000, China,\\
$^{2}$Institute of Theoretical Physics $\&$ Research Center of Gravitation, Lanzhou University, Lanzhou 730000, People's Republic of China}

\begin{abstract}
Recent explorations of topological aspects in black hole thermodynamics have achieved unprecedented progress. By utilizing topological numbers, different black hole systems can be categorized into distinct universality classes. This universal classification is particularly evident in thermodynamic limits, offering valuable insights for developing a comprehensive quantum gravity framework. This review highlights the latest advancements in this field. Specifically, we outline fundamental topological frameworks underlying black hole solutions, critical points, Davies points, and the Hawking-Page phase transition. For each scenario, we calculate the associated topological numbers and analyze their physical significance. Furthermore, we explore the practical implications arising from this research.
\end{abstract}



\maketitle

\section{Introduction}
\label{secIntroduction}

A black hole represents a region of spacetime where gravitational force is so immense that even light cannot escape. This phenomenon arises when the escape velocity at the event horizon surpasses the speed of light. From this fact, it is seems that there is no concept of temperature for black holes, and they are just pure gravity object. However, a pivotal thought experiment involving pouring hot tea into a black hole raised intriguing questions, as the act seemed to erase the entropy of the tea, challenging the second law of thermodynamics \cite{Bekenstein}.

To reconcile this inconsistency, Bekenstein proposed that black holes possess thermodynamic entropy, directly tied to the area of their event horizons \cite{Bekensteina,Bekensteinb}. Initially skeptical of this notion, Hawking's skepticism waned upon considering quantum effects near the black hole's event horizon, leading to the revelation that black holes emit particles outward at a characteristic temperature, now recognized as the Hawking temperature \cite{Hawking,Hawkingb}. Consequently, it was established that the entropy of a black hole equates to one-quarter of its horizon area, marking a crucial milestone in our understanding of these enigmatic cosmic entities \cite{Bekensteinb,Hawking}.

Following the seminal contributions of Bekenstein and Hawking, the framework of black hole thermodynamics has been formalized into four fundamental laws \cite{Bardeenab}. This formalization essentially maps a system of strong gravitational force onto a thermodynamic framework, bridging the gap between these disparate domains. Notably, the incorporation of quantum effects into black hole thermodynamics presents a crucial avenue for scrutinizing these two conceptually distinct paradigms, offering valuable insights into their intersection and potential compatibility.

Among the notable properties within thermodynamic systems, Hawking and Page uncovered a remarkable phase transition from a pure radiation phase to a stable large Schwarzschild black hole in Anti-de Sitter (AdS) space  \cite{Page}. This transition was used to interpreted the confinement/deconfinement phase transition in gauge field theories \cite{Witten2}.

Motivated by the AdS/CFT correspondence \cite{Maldacena,Gubser,Witten}, the study of black hole thermodynamics and phase transitions has garnered significant interest within the extended phase space, where the cosmological constant is interpreted as a form of pressure \cite{Kastor}. This framework has unveiled diverse types of phase transitions and intricate phase structures warranting further exploration \cite{Kubiznak,Altamirano,AltamiranoKubiznak,Altamirano3,Dolan,Liu,Wei2,Frassino,Cai,XuZhao,Hennigar,Hennigar2,Tjoa2,Teo,xuuuu3,Yangangang5,MMMMa2,Mann:2025xrb}.

Topology has emerged as a robust tool for probing the intrinsic properties of black holes. Utilizing the Brouwer degree of a continuous map, a recent study \cite{Cunhaa} demonstrated that light rings, crucial features associated with ultra-compact objects resulting from classical gravitational collapse under the null energy condition, invariably manifest in pairs in axisymmetric and stationary configurations. Notably, one of these pairs exhibits stability while the other remains unstable. This topological analysis, independent of specific spacetime dynamics, unveils a fundamental property inherent to spacetime structures.

Expanding their investigation to black hole backgrounds, the researchers established a pivotal theorem \cite{Cunhab}: in a (3+1)-dimensional asymptotically flat blackhole spacetime featuring a non-extremal and topologically spherical Killing horizon, each rotational direction hosts at least one standard (unstable) light ring beyond the horizon. This insight offers a broad perspective on the generation of black hole shadows, originating from the presence of unstable photon spheres or light rings, in non-spinning or spinning spacetimes, respectively.

Extending these analyses to asymptotic Anti-de Sitter (AdS) and de Sitter (dS) spacetimes has yielded consistent results \cite{Weiprd20}. Notably, the identification of paired timelike circular orbits with specific angular momentum in black hole spacetimes through analogous topological argument \cite{Wei:2022mzv} emphasizes the ability of the topological approach in shedding light on the nature of black holes that remains to be fully uncovered.

As previously highlighted, black holes showcase intricate thermodynamic phenomena and phase transition dynamics, prompting inquiries into potential thermodynamic universality. Drawing inspiration from topological investigations, we aim to establish a topological framework for black hole thermodynamics. This review delves into the rapid advancements and prospective applications in this field.

Our discussion commences with an exploration of Duan's $\phi$-mapping topological current theory, a pivotal tool for constructing the topology of black hole thermodynamics. Subsequently, we devote into topological methodologies concerning critical points, Davies points, Hawking-Page phase transition points, and the black holes themselves. We then shift focus towards delineating the topology associated with diverse black hole solutions, unraveling their unique topological attributes. Building upon these insights, we propose a universal topological classification scheme. Finally, we also discuss the potential applications of the black hole thermodynamics.

\section{Topological current and charge}
\label{tcac}

In this section, we present Duan's $\phi$-mapping topological current theory, designed to elucidate the topological characteristics associated with the zero points of a vector.

Let us start with a vector $\phi$ reformulated as
\begin{equation}
 \phi=||\phi||e^{i\Theta},\label{pp9}
\end{equation}
where $||\phi||=\sqrt{\phi^a\phi^a}$. Here, we mainly focus on the zeros of $\phi$. However, following (\ref{pp9}), its direction can not be uniquely defined at the zeros. In order to deal with it, one can express the vector as
\begin{equation}
 \phi=\phi^r+i\phi^\theta.
\end{equation}
Only when both $\phi^r$ and $\phi^\theta$ are zero, $\phi$ can attain 0. The unit vector is
\begin{equation}
 n^a=\frac{\phi^a}{||\phi||}, \quad a=1,2,\label{nonon}
\end{equation}
with $\phi^1=\phi^r$ and $\phi^2=\phi^\theta$. Note that the vector can have more than two components. In order to examine the topological charge, a superpotential is necessary
\begin{equation}
 V^{\mu\nu}=\frac{1}{2\pi}\epsilon^{\mu\nu\rho}\epsilon_{ab}n^{a}\partial_{\rho}n^{b},
 \quad\mu,\nu,\rho=0,1,2,
\end{equation}
where $x^{\mu}=(t$, $r$, $\theta$) and $\partial_{\mu}=\partial/\partial x^{\mu}$. Here, we take $t$ as a control parameter, which could be the time or other system parameter. From the expression, one easily obtain $V^{\mu\nu}=-V^{\nu\mu}$. Employing with the superpotential, the topological current can be defined as \cite{Duana,Duanb}
\begin{equation}
 j^{\mu}=\partial_{\nu}V^{\mu\nu}
  =\frac{1}{2\pi}\epsilon^{\mu\nu\rho}\epsilon_{ab}\partial_{\nu}n^{a}\partial_{\rho}n^{b}.
  \label{curr}
\end{equation}
Then it is easy to check
\begin{equation}
 \partial_{\mu}j^{\mu}=0,
\end{equation}
which implies that $j^{\mu}$ is a conserved current. Just like the vector of the electric current, the component $j^0$ denotes the charge density. Integrating it, we will obtain the topological charge corresponding the vector at given region $\Sigma$,
\begin{equation}
 \mathcal{Q}=\int_\Sigma j^{0}d^{2}x.
\end{equation}
Now, let us focus on the topological current $j^{\mu}$. Substituting (\ref{nonon}) into (\ref{curr}), one can obtain
\begin{equation}
 j^{\mu}=\frac{1}{2\pi}\epsilon^{\mu\nu\rho}\epsilon_{ab}\frac{\partial}{\partial\phi^{c}}
 \left(\frac{\phi^a}{||\phi||^2}\right)\partial_{\nu}\phi^{c}\partial_{\rho}\phi^{b}.
\end{equation}
Further making use of $\frac{\partial \ln||\phi||}{\partial\phi^a}=\frac{\phi^a}{||\phi||^2}$, it can be expressed as
\begin{equation}
 j^{\mu}=\frac{1}{2\pi}\epsilon^{\mu\nu\rho}\epsilon_{ab}
 \left(\frac{\partial}{\partial\phi^{c}}\frac{\partial}{\partial\phi^{a}}\ln||\phi||\right)
 \partial_{\nu}\phi^{c}\partial_{\rho}\phi^{b}.
\end{equation}
Considering the Jacobi tensor
\begin{equation}
 \epsilon^{ab}J^{\mu}\left(\frac{\phi}{x}\right)=\epsilon^{\mu\nu\rho}
 \partial_{\nu}\phi^a\partial_{\rho}\phi^b,\label{jaco}
\end{equation}
we have
\begin{equation}
 j^{\mu}=\frac{1}{2\pi}\left(\Delta_{\phi^a}\ln||\phi||\right)J^{\mu}\left(\frac{\phi}{x}\right),
\end{equation}
where $\Delta_{\phi^a}=\frac{\partial}{\partial\phi^a}\frac{\partial}{\partial\phi^a}$. In $\phi$-mapping space, the two-dimensional Laplacian Green function is
\begin{equation}
 \Delta_{\phi^a}\ln||\phi||=2\pi\delta(\phi).
\end{equation}
Therefore, the topological current will be in the following form
\begin{equation}
 j^{\mu}=\delta^{2}(\phi)J^{\mu}\left(\frac{\phi}{x}\right).\label{juu}
\end{equation}
From this expression, we are clear that $j^{\mu}$ is only nonzero at the zero points of $\phi^{a}$, i.e., $\phi^a(x^i, t)=0$. When the Jacobi determinant $J^{0}\left(\frac{\phi}{x}\right)\neq0$, one has $(\partial_{\mu}\phi^a)dx^{\mu}=0$, leading to the following result from (\ref{jaco})
\begin{equation}
 \epsilon_{\mu\nu\rho}J^{\mu}\left(\frac{\phi}{x}\right)dx^{\nu}=0.
\end{equation}
Multiplying both sides of the above equation by $\epsilon^{\lambda\sigma\rho}$, one has
\begin{equation}
 J^{\mu}\left(\frac{\phi}{x}\right)dx^{\nu}=J^{\nu}\left(\frac{\phi}{x}\right)dx^{\mu}.
\end{equation}
A simple calculation of $u^i=dx^i/dt$ gives
\begin{equation}
 J^{i}\left(\frac{\phi}{x}\right)=u^i J^{0}\left(\frac{\phi}{x}\right).\label{uxt}
\end{equation}
As a result, the components of $j^{\mu}$ are
\begin{eqnarray}
 j^{i}&=&\delta^{2}(\phi)J^{0}\left(\frac{\phi}{x}\right)u^i,\\
 j^{0}&=&\delta^{2}(\phi)J^{0}\left(\frac{\phi}{x}\right).
\end{eqnarray}
Therefore, the topological charge reads
\begin{eqnarray}
 \mathcal{Q}=\int_{\Sigma}\delta^{2}(\phi)J^{0}\left(\frac{\phi}{x}\right)d^2x.\label{qcharge}
\end{eqnarray}
Obviously, due to the $\delta$ function, the charge $\mathcal{Q}$ might be nonzero only if the region $\Sigma$ contains one or more zero points of the field. Thus we can assign each zero with a topological charge $\mathcal{Q}$.

Furthermore, considering there are $N$ zero points of $\phi$ and the Jacobi determinant $J^{0}\left(\frac{\phi}{x}\right)\neq0$, the solution of $\phi$=0 can be expressed as
\begin{equation}
 x^i=z^{i}_{n}(t), \quad n=1,2,...,N.
\end{equation}
Near the zero points, $\delta^{2}(\phi)$ can be expressed as the following form
\begin{equation}
 \delta^{2}(\phi)=\sum_{n=1}^{N}\alpha_{n}J^{0}\left(\frac{\phi}{x}\right)\bigg|_{x=z_n},
\end{equation}
where the coefficient $\alpha_{n}$ is positive.

According to Duan's topological current theory \cite{Duana,Duanb}, the winding number of the $n$-th zero point is
\begin{equation}
 w_{n}=w(\phi, z_n)=\alpha_n J^0\left(\frac{\phi}{x}\right)\big|_{x=z_n}.
\end{equation}
Considering $\alpha_n$ is positive, we have
\begin{equation}
 \alpha_n=\frac{|w(\phi, z_n)|}{|J^0\left(\frac{\phi}{x}\right)\big|_{x=z_n}}.
\end{equation}
At the zero point $z_n$, the $\phi$-mapping Hopf index $\beta_n$ and the Brouwer degree $\eta_n$ are given by
\begin{equation}
 \beta_n=|w(\phi, z_n)|,\quad
 \eta_n=\frac{J^{0}\left(\frac{\phi}{x}\right)}{|J^{0}\left(\frac{\phi}{x}\right)|_{x=z_n}}.
\end{equation}
As a result, we reach
\begin{equation}
 J^{0}\left(\frac{\phi}{x}\right)\delta^2(\phi)=\sum_{n=1}^{N}\beta_n\eta_n\delta^2(x-z_n).
\end{equation}
Finally, the topological charge can be expressed as
\begin{equation}
 \mathcal{Q}=\sum_{n=1}^{N}w_n=\sum_{n=1}^{N}\beta_n\eta_n.\label{windnu}
\end{equation}
This relation reflects the inner topological structure of the charge.

As demonstrated earlier, we have assumed $J^{0}\left(\frac{\phi}{x}\right)\neq0$. If this condition is violated, phenomena such as generation and annihilation may arise. To illustrate this, let us consider a situation where at least one component of the Jacobi tensor, for instance $J^{1}\left(\frac{\phi}{x}\right)$, does not equal zero. Consequently, one may observe
\begin{equation}
 \frac{dx^1}{dt}\bigg|_{(t_*,z_n)}=\frac{J^{1}\left(\frac{\phi}{x}\right)}
    {J^{0}\left(\frac{\phi}{x}\right)}\bigg|_{(t_*,z_n)}=\infty,
\end{equation}
which leads to
\begin{equation}
 \frac{dt}{dx^1}|_{(t_*,z_n)}=0.
\end{equation}
Then near the critical point ($t_*$, $z_n$), we have the following Taylor expansion
\begin{equation}
 t-t_*=\frac{1}{2}\frac{d^2t}{d(x^1)^2}\big|_{(t_*,z_n)}(x^1-z^1_n)^2.
\end{equation}
If $\frac{d^2t}{d(x^1)^2}|{(t*, z_n)}$ is less (greater) than zero, it signifies annihilation (generation), as discussed by Fu et al. \cite{Fu}. Given the topological current's conservation, it follows that these two cases must possess opposite winding numbers.

Before closing this section, we would like to provide some remarks. From Eq. (\ref{windnu}), it is apparent that the topological charge $\mathcal{Q}$ takes on integer values. For convenience, we refer to this topological charge as the topological number $W$. By further combining Eq. (\ref{qcharge}), the topological number can be represented as:
\begin{equation}
 W=\frac{1}{2\pi}\oint_{\partial\Sigma} d\Omega=\sum_{n=1}^{N}w_n.
\end{equation}
where $\Omega=\arctan(\phi^\theta/\phi^r)$ represents the direction of the vector. $\partial\Sigma$ denotes the one-dimensional boundary of $\Sigma$, which forms a closed contour. Therefore, the topological number signifies the count of loops traced by $\phi^a$ in the vector space as $x^{\mu}$ orbits the zero point $z_n$.

For each zero point, a closed contour can be constructed that exclusively encloses the zero point, enabling the evaluation of the winding number, which remains independent of the specific loops employed. Hence, the winding number serves as a local topological charge for the zero points, with different types of zeros corresponding to distinct values of the winding number.

Alternatively, by considering large regions encompassing all possible areas within the studied systems, it is possible to derive the total topological number $W$ that characterizes the system. Different systems exhibit distinct values of $W$, allowing for the classification of these systems based on $W$. This distinction highlights global topological properties.

In conclusion, the examination of a system's topological properties can be approached from either a global or local perspective using the winding number $w$ or the topological number $W$.

\section{Topological approaches}
\label{Topological}

In this section, we aim to introduce various topological approaches designed to reveal distinct thermodynamic properties.

\subsection{Topology of critical point}

The topological analysis of the critical point is the first approach in exploring black hole thermodynamics \cite{WeiLiu}. This analysis categorizes the critical point into two distinct classes: the conventional and the novel. This investigation unveils unique classifications of thermodynamic properties specific to critical points.

Let us start with the first law of the black hole in the extended phase space
\begin{eqnarray}
 dM=TdS+VdP+\sum_{i}Y_i dx^{i},\label{fila}
\end{eqnarray}
where $P$ is associated with the negative cosmological constant in AdS space, while $Y_i dx^{i}$ represents the $i$-th chemical potential term, which varies depending on the specific black hole system under consideration. Notably, in the extended phase space, the black hole mass $M$ is reinterpreted as enthalpy rather than internal energy. By performing a Legendre transformation $G=M-TS$, the Gibbs free energy can be derived. Subsequently, employing the Maxwell equal area law on the isothermal curve or observing the swallowtail behavior of the Gibbs free energy facilitates the determination of first-order small-large black hole phase transitions. Generally, the coexistence curve ends at a critical point, indicating a second-order phase transition, which can be resolved by \cite{Kubiznak}
\begin{eqnarray}
 (\partial_{V}P)=(\partial_{V,V}P)=0.
\end{eqnarray}
Alternatively, from the differential form of the Gibbs free energy $dG=-SdT+VdP+\sum_{i}Y_i dx^{i}$, the critical point can also be determined by \cite{WeiChengLiu}
\begin{eqnarray}
 (\partial_{S}T)=(\partial_{S,S}T)=0.
\end{eqnarray}
To delineate the topology concerning the critical point, we will consider the latter scenario. As depicted in Eq. (\ref{fila}), the temperature can be represented as a function of entropy $S$, pressure $P$, and additional parameters $x^{i}$
\begin{eqnarray}
 T=T(S, P, x^{i}).\label{tptp}
\end{eqnarray}
By requiring $(\partial_{S}T)=0$, we can eliminate one parameter, for example the pressure, in (\ref{tptp}). Then we label the new thermodynamic potential with $\Phi$:
\begin{eqnarray}
 \Phi=\frac{1}{\sin\theta}T(S, x^{i}),\label{phinew}
\end{eqnarray}
where the factor $\frac{1}{\sin\theta}$ is an auxiliary factor with $\theta\in (0, \pi)$. Employing with the function $\Phi$, we can introduce a new vector field $\phi=(\phi^S, \phi^\theta)$  \cite{WeiLiu}:
\begin{eqnarray}
 \phi^S=(\partial_{S}\Phi)_{\theta,x^{i}},\quad
 \phi^\theta=(\partial_{\theta}\Phi)_{S,x^{i}}.\label{veccon}
\end{eqnarray}
Within the framework of Einstein gravity, the entropy $S$ of $d-$dimensional black holes obeys the proportionality $S\propto r_{h}^{d-2}$, in accordance with the Hawking-Bekenstein entropy-area law. For this reason, the horizon radius $r_h$ is adopted as a substitute for the entropy $S$ in a number of research papers. While the vector undergoes modification, its zero point remains unaltered, thus leaving the topological approach unaffected. However, if the entropy is not a monotonically increasing function, i.e., the extremal points $\partial S/\partial r_h=0$ exist in modified gravity theories, we must adopt the entropy $S$ rather than the horizon radius $r_h$. Note that the incorporation of the $\theta$ factor results in a two-component vector, rendering the study of the thermodynamic topology of black holes more intuitive.

Several aspects regarding the $\theta$ factor are worth discussing: 1) It originates from the topology of the light ring \cite{Cunhab}; 2) The values $\theta=0$ and $\pi$ serve as boundaries in the $\theta$ direction within the parameter space. At these boundaries, the vector $\phi$ introduced becomes perpendicular, facilitating the calculation of the topological number; 3) The zeros of the vector consistently occur at $\theta=\pi/2$ without any additional zeros introduced. It is important to mention that the sign of Eq. (\ref{phinew}) could potentially alter the sign of the winding number; however, for relative value considerations, our results remain unaffected.

As a convention, it designates the conventional critical point with $w=-1$ and the novel critical point with $w=1$ \cite{WeiChengLiu}. To elucidate the specific topological properties, we will utilize the charged Reissner-Nordstr\"{o}m (RN)-AdS black hole and the charged Born-Infeld (BI)-AdS black hole as exemplary cases.

\begin{figure}
\includegraphics[width=7cm]{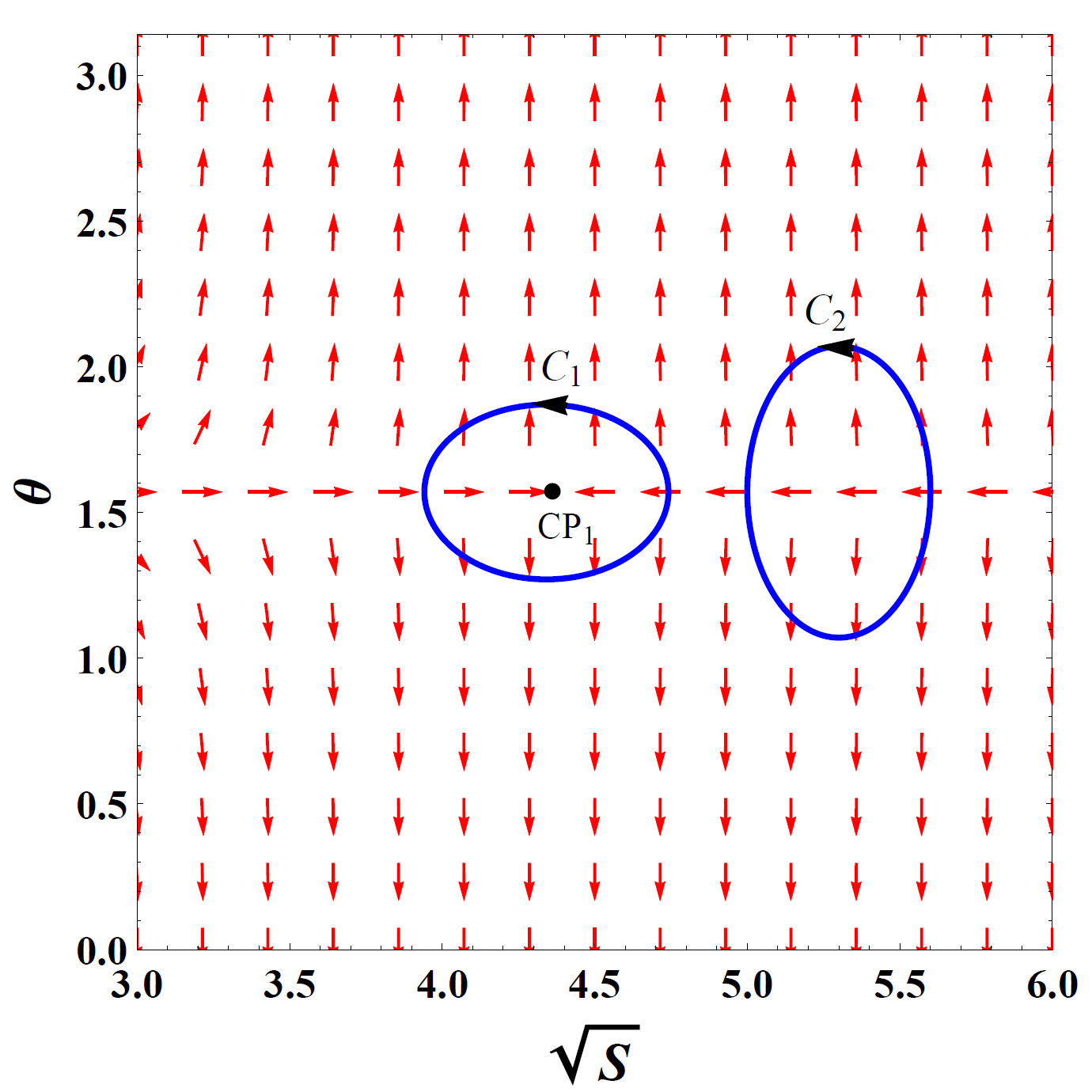}
\caption{The red arrows represent the vector field $n$ on a
portion of the $\sqrt{S}$-$\theta$ plane for the charged AdS black hole with the charge $q$=1. The critical point CP$_1$ located at ($\sqrt{S}$, $\theta$)=($\sqrt{6\pi}q$, $\frac{\pi}{2}$) is marked with a black dot. The blue contours $C_1$ and $C_2$ are two closed loops and $C_1$ encloses the critical point, while $C_2$ does not \cite{WeiLiu}.}\label{pCharVecss}
\end{figure}

For the charged RN AdS black hole, its Hawking temperature reads
\begin{eqnarray}
 T=\frac{2 P \sqrt{S}}{\sqrt{\pi }}-\frac{\sqrt{\pi } q^2}{4
   S^{\frac{3}{2}}}+\frac{1}{4 \sqrt{\pi } \sqrt{S}},\label{iscop}
\end{eqnarray}
where $q$ and $P$ are the charge and pressure of the black hole system, respectively. It is easy to check the first law
\begin{eqnarray}
 dM=TdS+\varphi dq+VdP
\end{eqnarray}
with $\varphi$ and $V$ being the electric potential and thermodynamic volume of the black hole system. Following the above general approach, we obtain the explicit form of the thermodynamic function
\begin{eqnarray}
 \Phi=\frac{1}{\sin\theta}\left(\frac{1}{2 \sqrt{\pi S }}-\frac{\sqrt{\pi }
   q^2}{S^{\frac{3}{2}}}\right),
\end{eqnarray}
as well as the components of the vector field $\phi$
\begin{eqnarray}
 \phi^{S}&=&\frac{\csc\theta \left(6\pi q^2-S\right)}{4 \sqrt{\pi } S^{\frac{5}{2}}},\\
 \phi^{\theta}&=&-\frac{\cot \theta \csc \theta \left(S-2 \pi q^2\right)}{2 \sqrt{\pi } S^{\frac{3}{2}}}.
\end{eqnarray}
The unit vector field $n=(\frac{\phi^{S}}{||\phi||}, \frac{\phi^{\theta}}{||\phi||})$ is shown in Fig. \ref{pCharVecss}. Obviously,  a critical point locates at $(\sqrt{S},\theta)=(\sqrt{6\pi}q,\frac{\pi}{2})$.

\begin{figure}
\includegraphics[width=7cm]{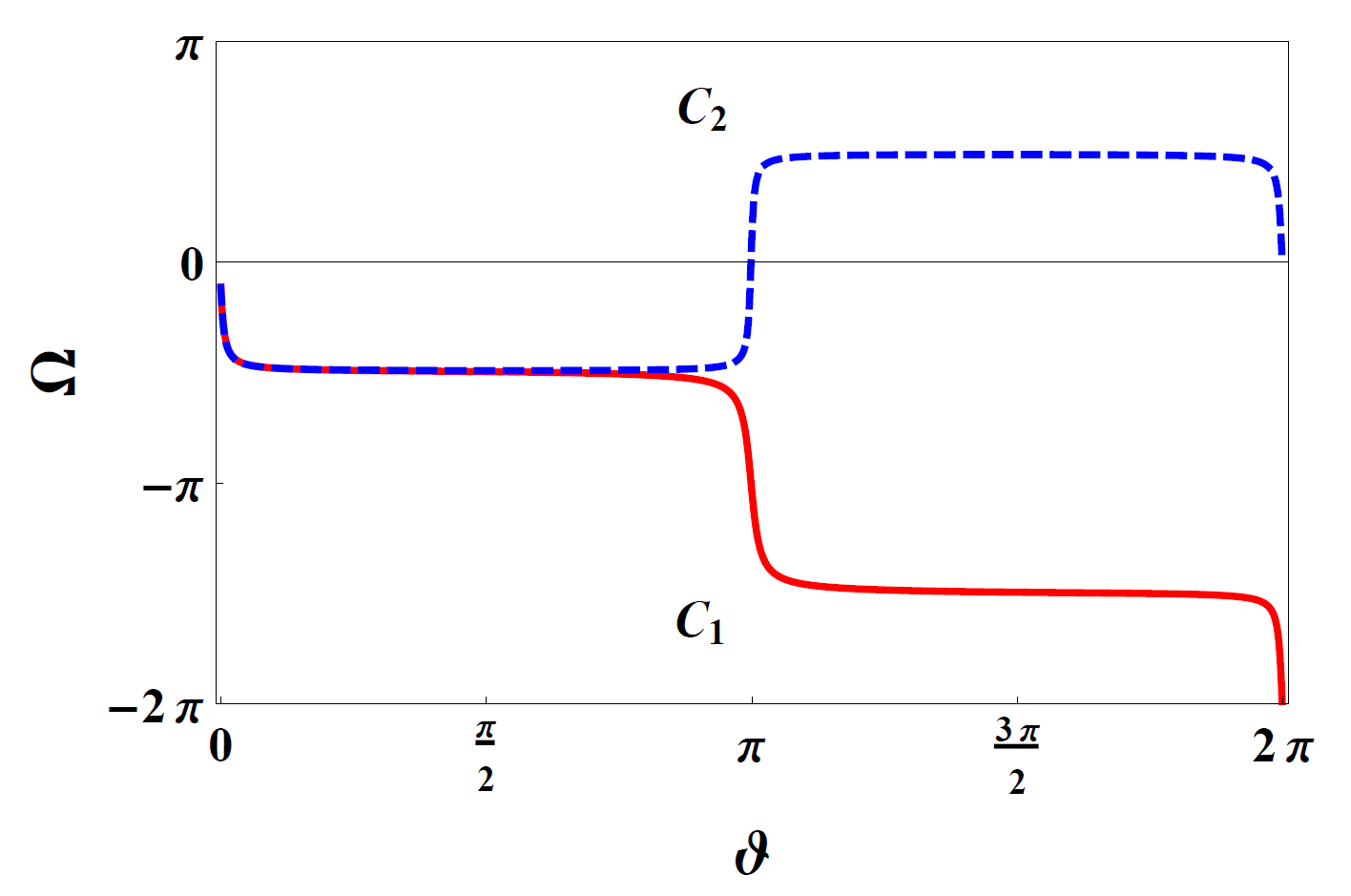}
\caption{$\Omega$ vs $\vartheta$ for the contours $C_1$ (red solid curve) and $C_2$ (blue dashed curve) \cite{WeiLiu}.}\label{pOmega}
\end{figure}

When the contour encompasses the critical point, it results in a non-zero winding; otherwise, the winding is zero. To determine the winding number, it is possible to create two contours $C_1$ and $C_2$, which are parameterized according to the following general form \cite{WeiLiu}:
\begin{eqnarray}
\left\{
\begin{aligned}
 r&=a\cos\vartheta+r_0, \\
 \theta&=b\sin\vartheta+\frac{\pi}{2},
\end{aligned}
\right.\label{pfs}
\end{eqnarray}
where $\vartheta$ ranges from 0 to $2\pi$ with the corresponding contours considered counterclockwise as positive. For $C_1$ and $C_2$, the values $(a, b, r_0)$ are set as (0.4, 0.3, $\sqrt{6\pi}$) and (0.3, 0.5, 5.3). To numerically compute the winding number,  a quantity measuring the deflection of the vector field along the specified contour is introduced
\begin{eqnarray}
 \Omega(\vartheta)=\int_{0}^{\vartheta}\epsilon_{ab}n^{a}\partial_{\vartheta}n^{b}d\vartheta.
\end{eqnarray}
Therefore, the winding number reads
\begin{eqnarray}
 w=\frac{1}{2\pi}\Omega(2\pi).
\end{eqnarray}
The behavior of $\Omega(\vartheta)$ for the contours $C_1$ and $C_2$ is illustrated in Fig. \ref{pOmega}. As $C_2$ does not encircle the critical point, we observe $w=0$ as expected. The plot distinctly demonstrates that with increasing $\vartheta$, $\Omega(\vartheta)$ initially decreases, then increases, ultimately converges to zero at $\vartheta=2\pi$, validating the anticipated outcome of $w=0$. In contrast, for $C_1$, $\Omega(\vartheta)$ progressively diminishes, approaching $2\pi$ at $\vartheta=2\pi$, resulting in a winding number of $w=-1$. According to our classification, this critical point is conventional. Notably, in its vicinity, a stable first-order small-to-large black hole phase transition transpires. Furthermore, as only one critical point is present, the topological number is
\begin{eqnarray}
 W=-1.
\end{eqnarray}

\begin{figure}
\includegraphics[width=7cm]{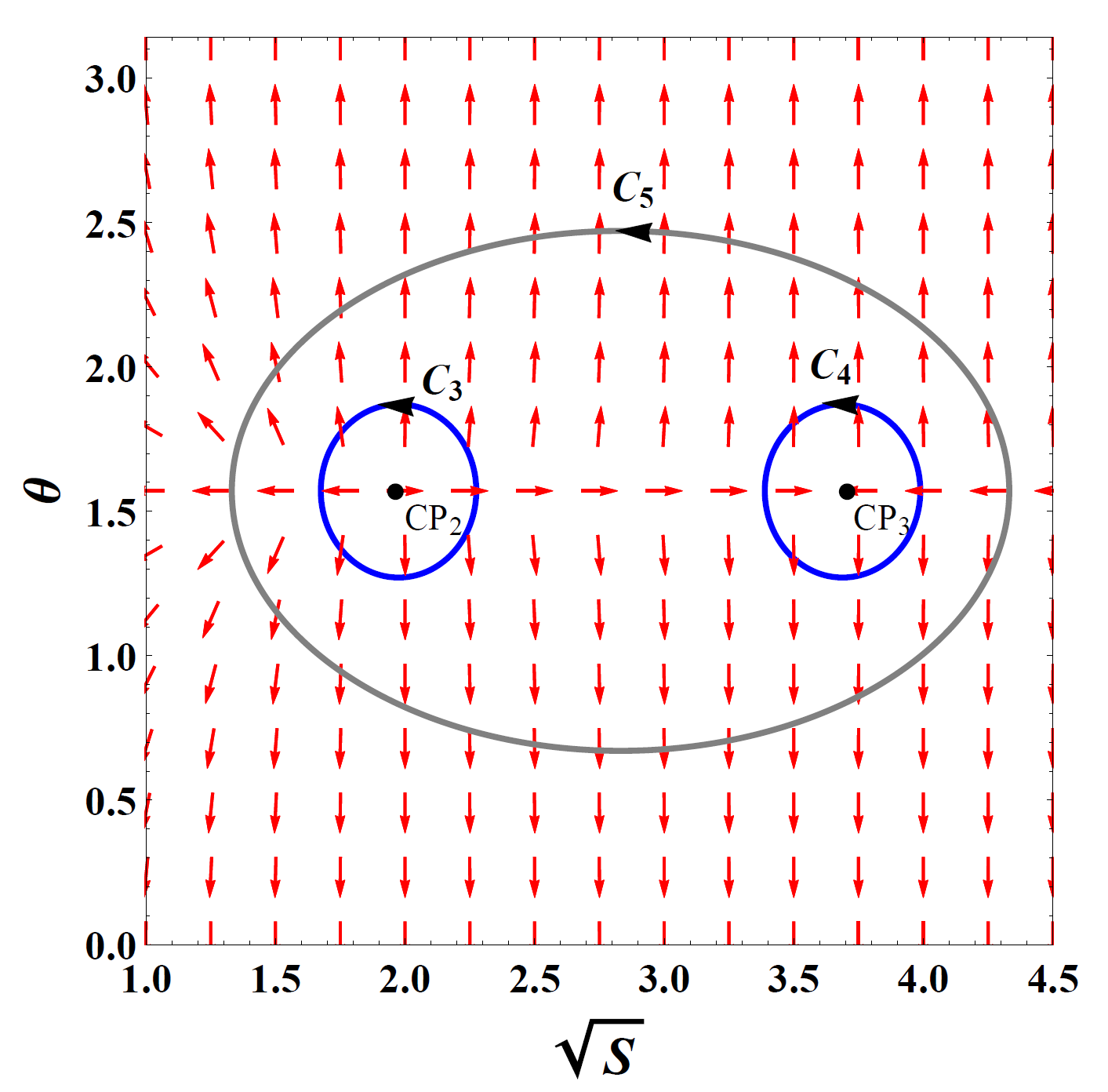}
\caption{The red arrows represent the vector field $n$ on a
portion of the $\sqrt{S}$-$\theta$ plane for the charged BI-AdS black hole with the charge $q$=1 and $b$=0.4. The critical points CP$_2$ and CP$_3$ located at ($\sqrt{S}$, $\theta$)=(1.97, $\frac{\pi}{2}$) and (3.69, $\frac{\pi}{2}$) are marked with black dots, and they are enclosed with the blue contours $C_3$ and $C_4$, respectively, while the gray contour $C_5$ encloses both the critical points \cite{WeiLiu}.}\label{pBIVec}
\end{figure}

\begin{figure}
\includegraphics[width=7cm]{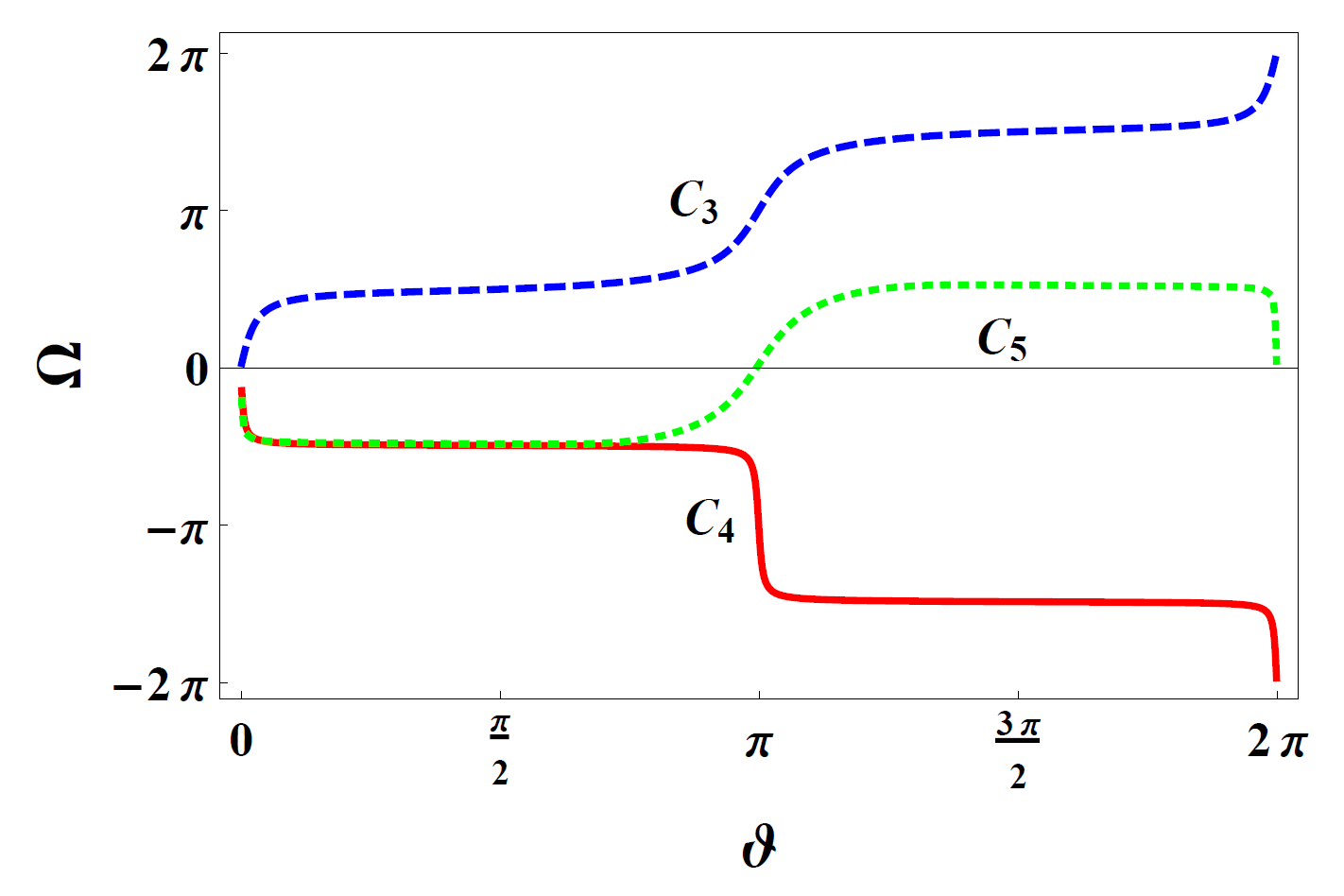}
\caption{$\Omega$ vs $\vartheta$ for the contours $C_3$ (solid curve), $C_4$ (dashed curve), and $C_5$ ( dot dashed curve) for the charged BI-AdS black hole \cite{WeiLiu}.}\label{pBICh}
\end{figure}

In order to exhibit the novel critical point, the BI-AdS black hole was considered \cite{Fernando}. Its Hawking temperature takes the following form \cite{Gunasekaranad}
\begin{eqnarray}
 T=\frac{1}{4\sqrt{\pi^3 S}}\left(2 \tilde{b}^2 S-2 \sqrt{\tilde{b}^4 S^2+\pi ^2 \tilde{b}^2 q^2}+8 \pi  P S+\pi\right),
\end{eqnarray}
where $\tilde{b}$ denotes the maximal electromagnetic field strength. The corresponding thermodynamic function is
\begin{eqnarray}
 \Phi=\frac{1}{2\sqrt{\pi S}\sin\theta}\left(1-\frac{2\pi \tilde{b} q^2}{\sqrt{\tilde{b}^2 S^2+\pi ^2 q^2}}\right).
\end{eqnarray}
The vector field $\phi$ and the normalized vector field $n$ can be readily obtained, allowing us to observe the behavior of the normalized vector field $n$ as depicted in Fig. \ref{pBIVec} with $q=1$ and $\tilde{b}=0.4$ for the charged BI-AdS black hole. In this particular scenario, two critical points, CP$_2$ and CP$_3$, are identified. Moreover, three contours, $C_3$, $C_4$, and $C_5$, are constructed with parameterized expressions akin to (\ref{pfs}), albeit with distinct parameter sets $(a, b, r_0)$=(0.3, 0.3, 1.97), (0.3, 0.3, 3.69), and (1.5, 0.9, 2.83), respectively. The deflection angle $\Omega(\vartheta)$ is calculated for these contours as shown in Fig. \ref{pBICh}, revealing diverse trends. Along $C_3$, $\Omega(\vartheta)$ ascends, descends along $C_4$, and exhibits an initial decrease followed by an increase and then a subsequent decrease along $C_5$. The values of $\Omega(2\pi)$ for these contours are 2$\pi$, $-2\pi$, and 0, respectively. Consequently, the topological charges are $w_{\text{CP}2}=1$ and $w_{\text{CP}_3}=-1$ for the critical points CP$_2$ and CP$_3$, categorically distinct due to their varying values. CP$_3$ denotes the conventional critical point, whereas CP$_2$ represents a novel one. Noteworthy, the topological number for the charged BI-AdS black hole is
\begin{eqnarray}
 W=w_{\text{CP}_2}+w_{\text{CP}_3}=0,
\end{eqnarray}
which is equivalent to that along the contour $C_5$. Due to the distinct topological numbers of the charged RN-AdS black hole and BI-AdS black hole, they fall into different topological classes. This observation prompts a deeper exploration into classifying black hole systems based on their topology.

We now observe that the critical points of the phase transition are categorized into two distinct classes. The behaviors of the vector near these zero points exhibit notable discrepancies. Consequently, exploring any thermodynamic distinctions arising from these differences is also warranted.

\begin{figure}
\label{vecplot}\includegraphics[width=7cm]{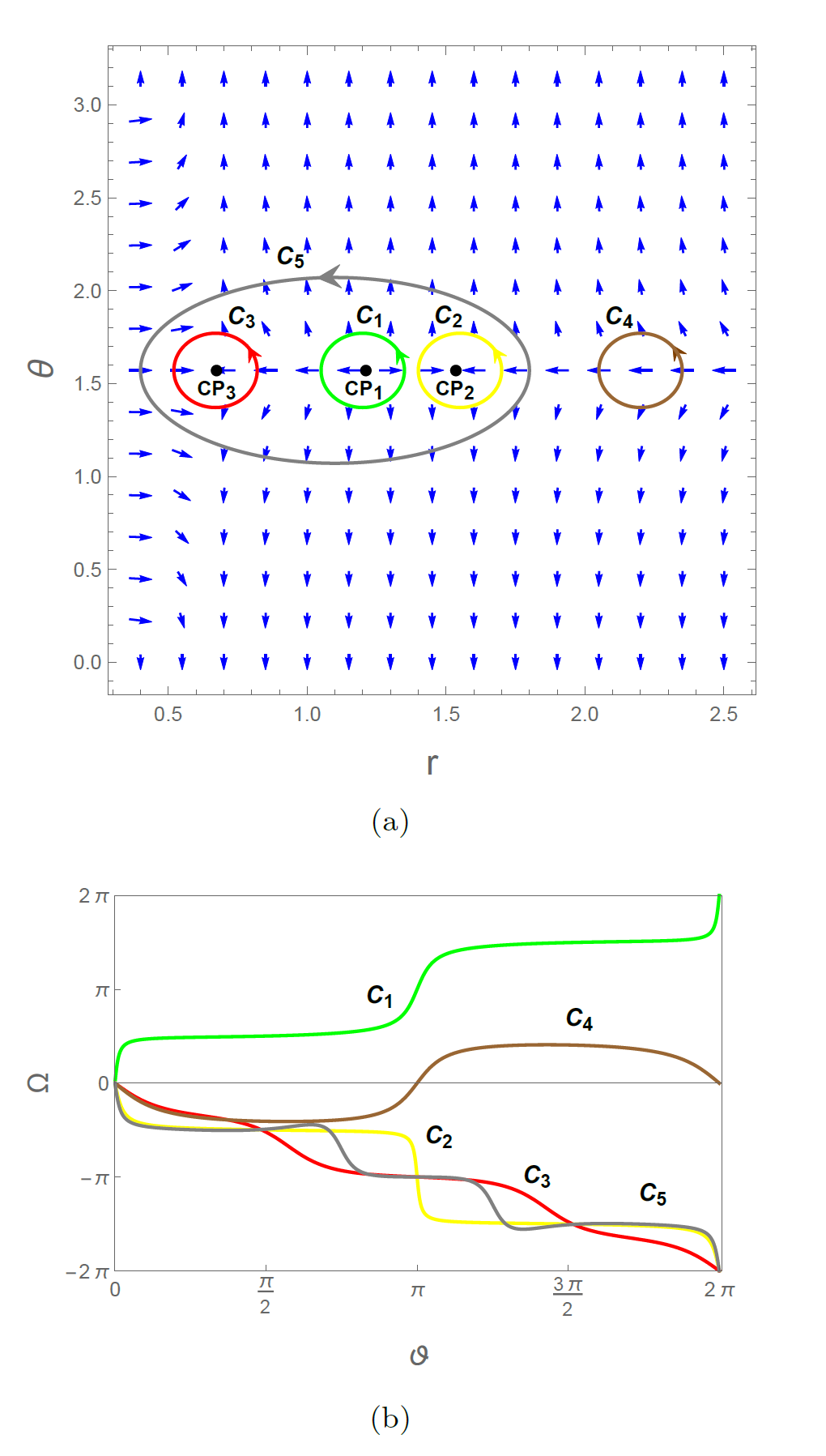}
\caption{(a) The blue arrows represent the vector field $n$ on a portion of the $\theta-r$ plane. The critical points $\text{CP}_1$, $\text{CP}_2$, and $\text{CP}_3$ are located at $(r,\theta)=(1.21, \frac{\pi}{2}), (1.53, \frac{\pi}{2}), \, \text{and} \, (0.67, \frac{\pi}{2}) $ marked with  black dots, and they are enclosed with the contours $C_1$, $C_2$ and $C_3$, respectively.
			The  contour $C_4$  does not enclose any critical point, while the contour $C_5$ encloses all the three critical points. (b) $\Omega$ vs $\vartheta$ for the contours $C_1$ (green curve),  $C_2$ (yellow curve), $C_3$ (red curve), $C_4$ (brown curve), and $C_5$ (grey curve) \cite{Yerra}.}\label{Ppomegaplot}
\end{figure}

In Ref. \cite{Yerra}, a six-dimensional charged Gauss-Bonnet AdS black hole was studied. A triple point and multiple critical points have been observed \cite{Wei2}. The constructed vector reads \cite{Yerra}
\begin{eqnarray}
\phi^r &=& \frac{\text{csc}\theta \Big(2Q^2(7r_h^2+30\alpha) - 3r_h^4(r_h^2-2\alpha)^2\Big)}{2\pi r_h^6 (r_h^2+6\alpha)^2}, \\
\phi^\theta &=& -\frac{\text{cot}\theta \, \text{csc}\theta (3r_h^6+2r_h^4\alpha -2Q^2)}{2\pi r_h^5 (r_h^2+6\alpha)}.
\end{eqnarray}
Such a vector can have up to three zero points. By selecting $\alpha=1$ and $Q=0.15$, the unit vector of $\phi$ is illustrated in Fig. 5(a), clearly displaying three zero points. The parameter $\Omega$ is also depicted in Fig. 5(b). It is straightforward to derive $w_{C_1}=1$, $w_{C_2}=w_{C_3}=w_{C_5}=-1$, and $w_{C_4}=0$. For the system, the total topological number is
\begin{eqnarray}
 W=w_{C_1}+w_{C_2}+w_{C_3}=-1,
\end{eqnarray}
suggesting its topological equivalence to the charged RN-AdS black hole but differentiation from the BI-AdS black hole. Notably, through variations in $\alpha$ and $Q$, it was observed that the topological number remains constant. Moreover, in Lovelock gravity, the creation of Vortex/anti-vortex pairs was explored, revealing the conservation of the topological number \cite{Ahmed}.

By relinquishing the global stability of the free energy, a proposition emerges that the conventional critical point signifies the proximity of first-order phase transitions, whereas the novel critical point does not necessarily indicate the presence of such transitions \cite{WeiLiu}. Conversely, when accounting for global stability, certain critical points may vanish if they fail to globally minimize the Gibbs free energy. Addressing this concern, in Ref. \cite{Yerra}, it was suggested that with increasing pressure, the novel critical point marks the emergence of new phases (whether stable or unstable), while the conventional point denotes the vanishing of phases. Further investigations into novel Einstein-Gauss-Bonnet gravity, boundary matrix duals, and Lovelock gravity can be found in Refs. \cite{YerraBhamidipati,Yerra:2023hui,Zhang:2024mqh}.

At this point, we have effectively delineated the thermodynamic topology of critical points and expounded upon the physical implications of the two distinct classes of critical points identified by the winding number. Moreover, we underscore the prospect of system-wide classification by leveraging the total topological number.

\subsection{Topology of Davies phase transition}

As a specific phase transition, Davies proposed that a phase transition occurs at the point where the heat capacity diverges \cite{Davies77}, also referred to as the spinodal point. At this point, the discontinuity in the first derivative of free energy implies a second-order phase transition. This scenario differs significantly from a critical point where the neighboring phases are locally thermodynamically stable; for the Davies phase transition point, one neighboring phase is stable while the other is unstable. Despite this distinction, a new topology can still be constructed for it \cite{Bhattacharya:2024bjp,Mehmood:2024nqd,Hazarika:2024imk}.

The construction process parallels that of the critical point discussed in the previous subsection. The thermodynamic potential $\Phi$ is selected as described in \cite{Bhattacharya:2024bjp}
\begin{eqnarray}
 \Phi=\frac{1}{\sin\theta}\frac{1}{T},
\end{eqnarray}
rather than (\ref{phinew}).
Consequently, the vector reads
\begin{eqnarray}
 \phi^S&=&\partial_S \Phi(S,\theta)=\frac{1}{\sin\theta}\frac{\partial}{\partial S}\Big(\frac{1}{T}\Big)~,  \\
    \phi^{\theta}&=&\partial_{\theta}\Phi(S,\theta)=-\frac{\csc\theta\cot\theta}{T}.
\end{eqnarray}
One can easily find that the Davies points exactly locates at the zeros of the vector $\phi$. Therefore, we can calculate the winding number following the above subsection.

Here we could take the charged RN-dS black hole with $Q=1$ and $\Lambda=1$ as an example. After a simple calculation, the thermodynamic potential reads
 \begin{eqnarray}
     \Phi(S,\theta)=\frac{1}{\sin\theta}\frac{1}{T}\Big|_{Q=\eta=\Lambda=1}=\frac{4 \pi ^{3/2} S^{3/2}}{\sin\theta(-S^2+\pi  S-\pi ^2)}.
 \end{eqnarray}
Then the vector field $\phi$ can be obtained as \cite{Bhattacharya:2024bjp}
 \begin{eqnarray}
  \phi^S&=&-\frac{2 \pi ^{3/2} \sqrt{S} \left(-S^2-\pi  S+3 \pi ^2\right) \csc\theta}{\left(S^2-\pi  S+\pi ^2\right)^2},\\
 \phi^{\theta}&=&\frac{4 \pi ^{3/2} S^{3/2} \cot\theta \csc\theta}{S^2-\pi  S+\pi ^2}.
 \end{eqnarray}
The behavior of the unit vector of $\phi$ was examined in Ref. \cite{Bhattacharya:2024bjp}. Upon computation, the winding number for the Davies point is determined to be $w=-1$.

Consequently, the Davies point is afforded a topological interpretation. However, the physical implications of positive or negative winding numbers remain for further exploration.

\subsection{Topology of Hawking-Page phase transition}

The Hawking-Page phase transition signifies a transition between pure radiation and a large stable Schwarzschild black hole in AdS space \cite{Page}. This transition is prevalent across various AdS black hole systems. Pure radiation tends to be overshadowed by the black hole's parameters, hence manifesting solely in the grand canonical ensemble, where the chemical potential remains fixed as opposed to the black hole parameters.

The free energy reads
\begin{eqnarray}
 \mathcal{F}=M-\bar{T}S,
\end{eqnarray}
where $\bar{T}$ is the ensemble temperature rather than the Hawking temperature. Under this pattern, the Hawking-Page phase transition occurs exactly at \cite{Yerra:2022coh}
\begin{eqnarray}
 \mathcal{F}=0,\quad \frac{\partial \mathcal{F}}{\partial r_{h}}=0.
\end{eqnarray}
Solving the first equation, we will obtain $\bar{T}=(M-\mathcal{F})/S$. Then the Hawking-Page phase transition shall be determined by $\partial_{r_h}{\bar{T}}=0$. As a result, the thermodynamical potential is defined as
\begin{eqnarray}
 \Phi=\frac{\bar{T}}{\sin\theta}.
\end{eqnarray}
By utilizing this approach, the vector $\phi$ can be derived following Eq. (\ref{veccon}). The Hawking-Page phase transition point precisely corresponds to the zero of $\phi$, thus establishing the topology for the Hawking-Page phase transition. This analysis was extended to BI-AdS and Kerr-AdS black holes in Refs. \cite{Yerra:2023ocu,Hazarika:2024imk}.

\subsection{Topology of black hole solution}

As demonstrated earlier, we have established the topology for the critical point, Davies phase transition point, and Hawking-Page phase transition point. Here, we aim to present an additional thermodynamic topology concerning the black hole solution itself \cite{WeiLiuad2022}. This topology serves a dual purpose: it interprets local thermodynamic (in)stability as a topological property and offers a method for classifying black hole systems. Consequently, it has garnered significant attention and subsequent exploration. In this subsection, we intend to briefly introduce this topology.

The zeros of the vector can be viewed as defects in the thermodynamic parameter space, underscoring the significance of whether the black hole solution can be represented as zeros of the vector in constructing this topology. Drawing inspiration from this concept, one can reformulate the Einstein field equations as follows:
\begin{eqnarray}
 \mathcal{E}_{\mu\nu} \equiv G_{\mu\nu}-\frac{8\pi G}{c^4} T_{\mu\nu}=0.\label{pp}
\end{eqnarray}
It is evident that black hole solutions satisfying the Einstein field equations correspond to the zeros of $\mathcal{E}_{\mu\nu}$. This observation underscores the applicability of the defect theory to black hole solutions. The subsequent challenge entails identifying a suitable thermodynamic quantity to actualize this concept.

In the work \cite{WeiLiuad2022}, it was proposed that the generalized free energy introduced by York \cite{York1986} can serve this purpose. This free energy is designed to address the issue of local thermodynamic instability of the Schwarzschild black hole. By envisioning the black hole within a cavity, stability of a Schwarzschild black hole can be achieved if a specific relation between the black hole mass and the cavity temperature is met. In this scenario, mass and temperature are treated as independent variables. The corresponding free energy can be formulated as:
\begin{eqnarray}
 \mathcal{F}=M-\frac{S}{\tau},\label{grf}
\end{eqnarray}
where the parameter $\tau$ represents an additional variable with the dimensions of time and can be interpreted as the inverse temperature of the cavity containing the black hole. This free energy can also be referred to as the off-shell free energy. When
\begin{eqnarray}
\tau=T^{-1},\label{condd}
\end{eqnarray}
the black hole solution satisfies the Einstein field equations (\ref{pp}), thus representing an actual solution for a black hole.
Let's proceed by calculating the derivative of $\mathcal{F}$ with respect to the horizon radius $r_{h}$
\begin{eqnarray}
 \frac{\partial \mathcal{F}}{\partial r_{h}}&=&\frac{\partial E}{\partial r_{h}}-\frac{1}{\tau}\frac{\partial S}{\partial r_{h}}\nonumber\\
 &=&\frac{\partial E}{\partial S}\frac{\partial S}{\partial r_{h}}-\frac{1}{\tau}\frac{\partial S}{\partial r_{h}}\nonumber\\
 &=&\left(T-\frac{1}{\tau}\right)\frac{\partial S}{\partial r_{h}}.
\end{eqnarray}
In the third step, we have used the first law of black hole thermodynamics. Due to $\frac{\partial S}{\partial r_{h}}\neq0$, the on-shell condition translates to $\partial\mathcal{F}/\partial r_{h}=0$. Consequently, the vector can be defined as
\begin{eqnarray}
\phi=\left(\frac{\partial \mathcal{F}}{\partial r_{\text{h}}}, -\cot\Theta \csc\Theta\right). \label{vectorField}
\end{eqnarray}
At $\Theta=0$ and $\pi$, the component $\phi^{\Theta}$ exhibits a divergence, indicating an outward direction for the vector. Notably, the zeros of $\phi$ coincide with the conditions $\Theta=\pi/2$ and $\tau=T^{-1}$. This correspondence highlights the precise alignment of the black hole solution with the zero points of the vector $\phi$. Consequently, from a topological perspective, we can assign a winding number to each black hole solution by employing $\phi$.

Following by the Duan's $\phi$-mapping topological current theory, we can establish the topology for the black hole solution by using the constructed vector (\ref{vectorField}).

For the Schwarzschild black hole, the generalized free energy is given by $\mathcal{F}=\frac{r_{\text{h}}}{2}-\frac{\pi r_{\text{h}}^2}{\tau}$. Consequently, the vector $\phi$ can be expressed as
\begin{eqnarray}
\phi^{r_{\text{h}}}&=&\frac{1}{2}-\frac{2\pi r_{\text{h}}}{\tau},\\
\phi^{\Theta}&=&-\cot\Theta \csc\Theta.
\end{eqnarray}
We depict the unit vector field $n$ over a section of the $\Theta$-$r_{\text{h}}$ plane in Fig. 6(a), focusing on the Schwarzschild black hole with $\tau=4\pi r_0$, where $r_0$ represents an arbitrary length scale determined by the dimensions of a cavity enclosing the black hole.

\begin{figure}
\label{Topo}\includegraphics[width=5cm]{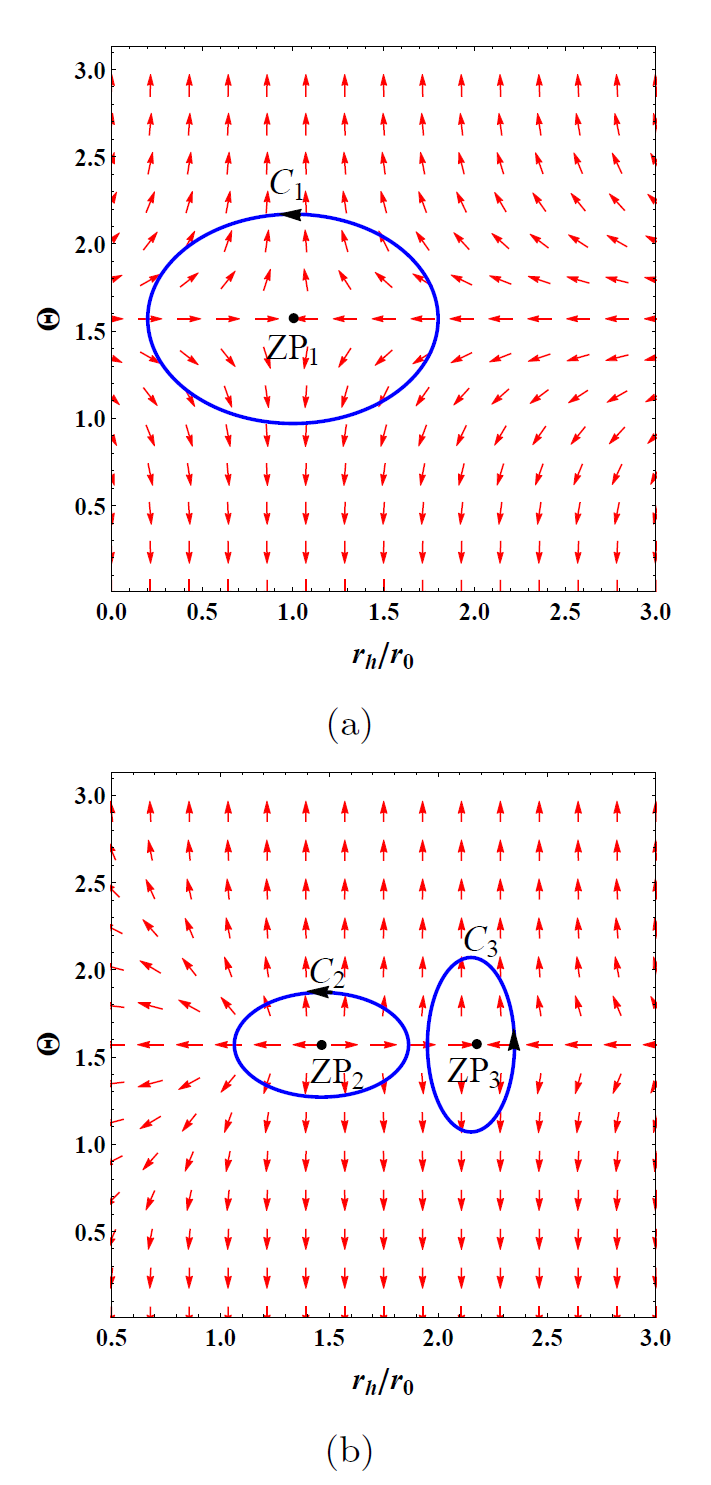}
\caption{The red arrows represent the unit vector field $n$ on a
portion of the $r_{\text{h}}$-$\Theta$ plane. The zero points (ZPs) marked with black dots are at ($r_{\text{h}}$, $\Theta$)=(1, $\frac{\pi}{2}$), (1.46, $\frac{\pi}{2}$), and (2.15, $\frac{\pi}{2}$), for ZP$_1$, ZP$_2$, and ZP$_3$, respectively. The blue contours $C_i$ are closed loops enclosing the zero points. (a) The unit vector field for the Schwarzschild black hole with $\tau/r_0=4\pi$. (b) The unit vector field for the RN black hole with $\tau/r_0=34.48$ and $Q/r_0=1$ \cite{WeiLiuad2022}.}\label{pCharVec}
\end{figure}

Examining the figure, the zero point is precisely located at $r_{\text{h}}/r_0=1$ and $\Theta=\pi/2$. Given that the winding number $w$ remains constant across contours enclosing this zero point, its calculation can be carried out for any arbitrary contour, such as $C_1$ illustrated in Fig. 6(a). Upon computation, the winding number is found to be $w=-1$. If an alternative orientation convention be adopted, involving a negative sign in $\phi^{r_{\text{h}}}$ within (\ref{vectorField}), the result could be altered to $w=1$. This adjustment would similarly impact the winding numbers for other types of black holes. Nonetheless, the underlying physical significance would remain unchanged.

For the RN black hole, the generalized free energy is given by:
\begin{eqnarray}
 \mathcal{F}=\frac{r_{\text{h}}^2+Q^2}{2 r_{\text{h}}}-\frac{\pi r_{\text{h}}^2}{\tau}.
\end{eqnarray}
By utilizing this expression, the vector can be readily derived. Subsequently, we present the unit vector field $n$ in Fig. 6(b) for $\tau/r_0=34.48$ and $Q/r_0=1$. We identify two zero points, ZP$_2$ and ZP$_3$, situated at $r_{\text{h}}/r_0=1.46$ and $2.15$ respectively. Through computations, their associated winding numbers are determined as $w=1$ and $-1$. Moreover, the heat capacities are calculated as $C_Q/r_{0}^2=18.01$ and $-64.81$ at a constant charge of $Q/r_0=1$ for the positive and negative zero points, respectively. Notably, each zero point of the unit vector has a winding number of $1$ or $-1$. This observation leads to a conjecture linking the winding number to local thermodynamic stability, where positive and negative values denote stable and unstable black hole solutions, respectively.

\begin{figure}
\includegraphics[width=7cm]{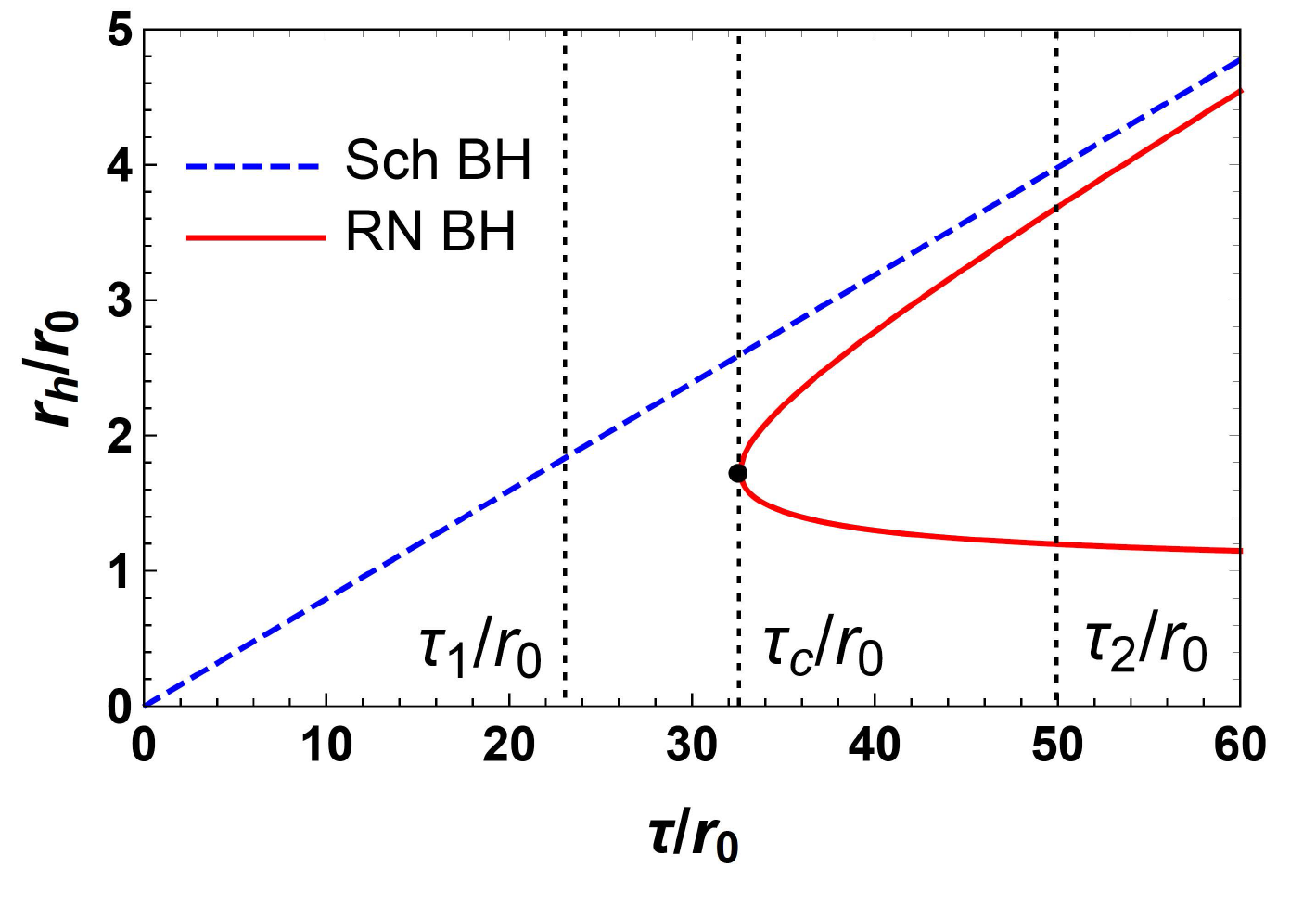}
\caption{Zero points of the vector $\phi$ shown in the $r_{\text{h}}$-$\tau$ plane. The blue dashed and red solid lines are for the Schwarzschild black hole (Sch BH) and RN black hole with $Q/r_0=1$. The black dot with $\tau_c=6\sqrt{3}\pi Q$ denotes the generation point for the RN black hole. At $\tau=\tau_1$, there is one Schwarzschild black hole, and at $\tau=\tau_2$, there are one Schwarzschild black hole and two RN black holes \cite{WeiLiuad2022}.}\label{rht}
\end{figure}

For the Schwarzschild and RN black holes, we determine the total topological numbers to be $W=-1$ and 0, respectively, indicating that they belong to distinct topological classes. This distinction can be readily discerned through the defect curve, which encompasses all zeros. To illustrate this, we present the defect curve for the Schwarzschild and RN black holes in Fig. \ref{rht}. In the scenario of large $\tau$ (e.g., $\tau=\tau_2$), the Schwarzschild and RN black holes display one and two intersection points, respectively. These points precisely satisfy the condition (\ref{condd}), denoting on-shell black hole solutions with a temperature of $T=\tau^{-1}$. Notably, unlike the Schwarzschild black hole, when $\tau$ decreases below $\tau_{c}$, the two intersection points of the RN black hole coalesce and subsequently disappear. This observation implies that variations in the parameter $\tau$ and black hole charge do not change the topological number.

\begin{figure}
\includegraphics[width=7cm]{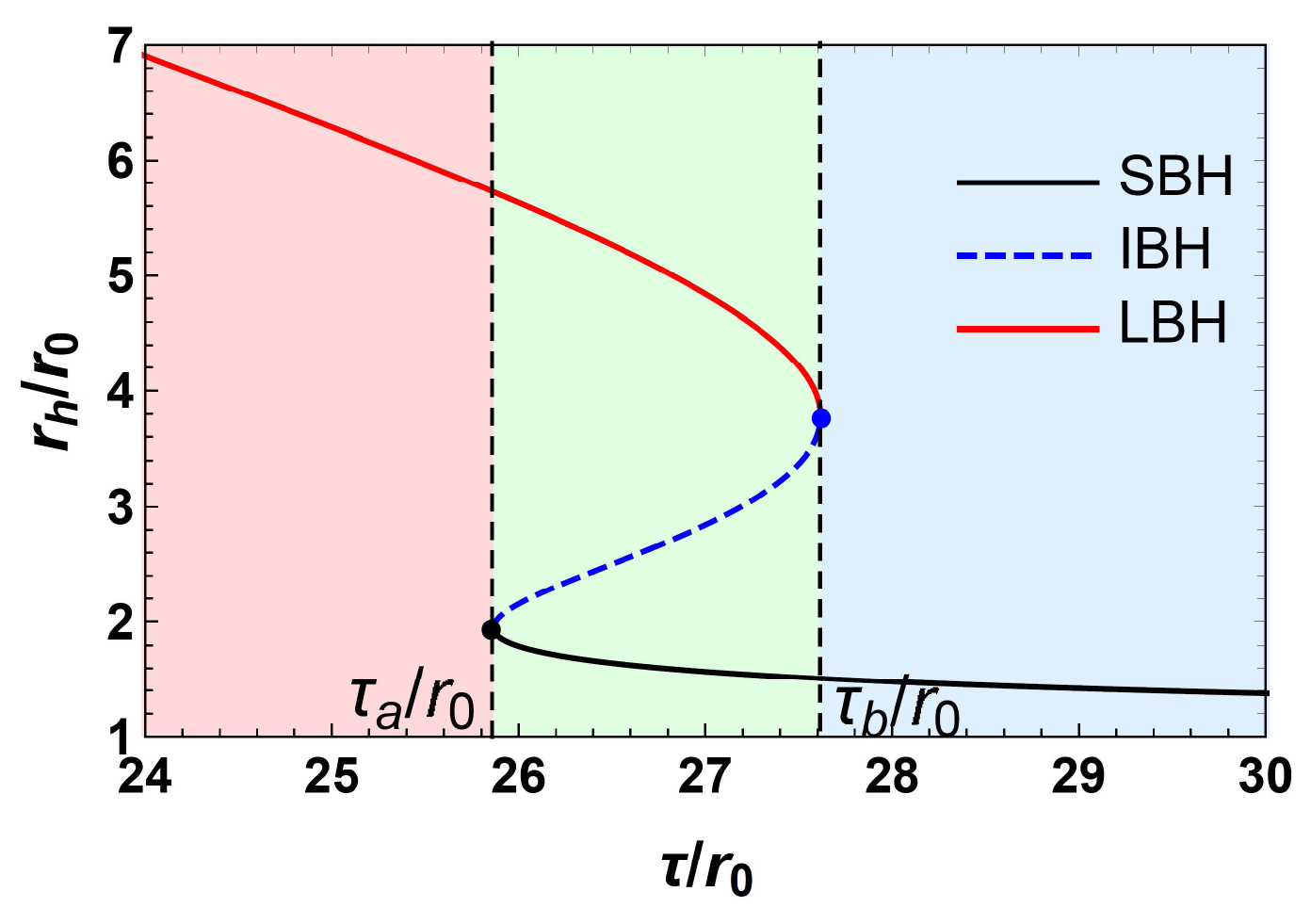}
\caption{Zero points of $\phi^{r_{\text{h}}}$ shown in the $r_{\text{h}}$-$\tau$ plane for the RN-AdS black hole with $Pr_{0}^{2}=0.0022$ and $Q/r_0=1$. The black solid, blue dashed, and red solid lines are for the small black hole (SBH), intermediate black hole (IBH), and large black hole (LBH), respectively. Black and blue dots are the annihilation and generation points. Different color regions have different number of the black hole branches. However their $W$ number is constant and equals 1 \cite{WeiLiuad2022}.}\label{rhAdS}
\end{figure}

Taking the charged RN-AdS black hole as another example, we investigate its characteristics. This black hole displays a small-large phase transition, potentially resulting in a distinct behavior of the defect curve. The results are depicted in Fig. \ref{rhAdS}. Unlike the Schwarzschild and RN black holes, a notable distinction arises with three distinct black hole branches observed for $\tau_{a}<\tau<\tau_{b}$: one small branch for $\tau<\tau_{a}$, one large branch for $\tau>\tau_{b}$, and an intermediate branch between them.

Upon computing the winding numbers for these branches, it is revealed that the small and large branches possess $w=1$, whereas the intermediate branch exhibits $w=-1$. Consequently, for the RN-AdS black hole under this pressure condition, the total winding number consistently computes to $W=1-1+1=1$, irrespective of $\tau$. Despite the positive pressure $P$ associated with the RN-AdS black hole, it does not impact the asymptotic behavior of $\tau$ at small and large $r_{\text{h}}$. Consequently, the topological number $W$ remains one for the RN-AdS black hole.

This subsection is dedicated to establishing the topology of the black hole solution. The winding number holds a unique physical significance, serving as a key indicator of the local thermodynamic stability of the black hole. A positive or negative winding number denotes the local stability or instability of the black hole.

It is noteworthy that for the Schwarzschild, RN, and charged RN-AdS black holes, their respective topological numbers are $W=-1$, 0, and 1, delineating their classification into distinct topological categories. Furthermore, the black hole parameters, such as charge and pressure, demonstrate no influence on the topological number, indicating a universal characteristic of the topology.

\subsection{Extended thermodynamical topology}

Based on the foregoing analysis, it is evident that diverse thermodynamic topologies have been constructed via the utilization of free energy. A critical question thus emerges: does there exist a universal theoretical framework that subsumes all the aforementioned topologies as its specific subcases?

Inspired by this idea, a $k-$th order vector field was defined in Ref. \cite{WeiYang2025}
\begin{equation}
 \Phi _k\left( S,\Theta \right)=\left( \left. \partial _{S}^{k+1}F \right|_{\left\{ \partial _{S}^{m}F=0,\ m=1,\ \dots,\ k. \right\}},-\cot \Theta \csc \Theta \right).
 \label{eq_Phikdef}
\end{equation}
The zero points correspond to
\begin{equation}
 \partial_S F = 0,\quad \partial_S^2 F = 0, \quad \dots\ , \quad \partial_S^{k+1} F = 0.
 \label{eq_PSkF}
\end{equation}
By setting $k=$1, 2, 3, the thermodynamic topologies corresponding to the black hole solution, Davies point, and critical point can be respectively recovered. For each zero point of the vector (\ref{eq_Phikdef}), the associated winding number is derivable through the analysis of the heat capacity $C$
\begin{equation}
 w^{(k)} = sign(\partial_S^k (C^{-1})).
\end{equation}
Then the extended thermodynamic topology is characterized by the following set of topological number
\begin{equation}
 \{W^{(k)}, k = 0, 1, 2, \dots\}.
\end{equation}
Furthermore, the scenario involving degenerate zero points has also been taken into account in Ref. \cite{WeiYang2025}. This provides a viable approach to characterize the thermodynamic topology corresponding to the isolated critical point within the framework of seven-dimensional Lovelock gravity.

In this section, we introduce several different thermodynamical topologies. The physical meanings of the winding number is also examined. For convenience, we summarize them in Table \ref{tab1}.

\begin{table*}[t]
\setlength{\tabcolsep}{3mm}{
\begin{center}
\begin{tabular}{cccc}
  \hline\hline
$w$ & -1 & 0 &1\\\hline
Critical point (CP) &conventional CP &degenerated CP/---&novel CP\\
Davies point &existence &degenerated point/--- &existence\\
Hawking-Page phase transition&---&---&existence\\
Black hole solution &unstable black hole&Davies point/---&stable black hole\\
Extended topology &-$\partial_S^k (C^{-1})$&degenerated point/---&$\partial_S^k (C^{-1})$\\
\hline\hline
\end{tabular}
\caption{Physical meanings of winding number for different thermodynamical black hole topologies. ``---'' denotes the trivial case or non-existence.}\label{tab1}
\end{center}}
\end{table*}

\section{Features of topological number}
\label{Features}

In the preceding section, we delineate four distinct topologies: the critical point, Davies point, Hawking-Page phase transition point, and the black hole solution itself. Among these, the fourth topology has garnered significantly more attention owing to its clear physical interpretation and universal applicability. Subsequent studies predominantly concentrate on this topology. Therefore, this section is dedicated to devote into the discussion on the topological number associated with this specific topology.

\subsection{Black hole parameters}

As demonstrated earlier, black hole charge and pressure do not exert an influence on the topological number for each certain black hole solution. Nonetheless, there remain other unexplored black hole parameters that warrant investigation.

As highlighted previously, black hole solutions with charge, cosmological constants, or pressures exhibit distinct topological numbers that remain invariant. An essential parameter yet to be examined is the angular momentum (or spin) of the black hole, in accordance with the no-hair theorem. Hence, exploring the topology of rotating black holes merits further research.

In the Refs. \cite{Wu:2022whe, Mohamed:2025iqn}, it was observed that by fixing the black hole spin $a/r_0$ (corresponding to angular momentum $J=a/M$) and charge $Q/r_0$, the $d=4$ and 5-dimensional Kerr black holes and Kerr-Newman black holes possess a topological number $W=0$, akin to the charged RN black hole without spin. However, for $d\geq6$, the singly rotating Kerr black hole demonstrates $W=-1$. This outcome underscores the significant impact of the dimensionality on the topological number for rotating black holes in asymptotically flat spacetime.

In asymptotic AdS spacetime, the $d=4$ and 5-dimensional Kerr-AdS, Kerr-Newman, and rotating BTZ black holes exhibit $W=1$, while for $d\geq6$, singly rotating Kerr-AdS black holes yield $W=0$ \cite{Wu:2023sue}. Conversely, for double-rotating AdS black holes, the topological number could be $W=-1$ \cite{Zhu:2024jhw}. These findings underscore that pressure or the cosmological constant can alter the topological number under certain different specific black hole solutions, yet the value of this number remains invariant within each individual case.

\subsection{C-metric and NUT spacetimes}

The presence of conical singularities near the south or north pole in C-metric and NUT spacetimes gives rise to intriguing phenomena around black holes \cite{Taub, Newman, Kinnersley}.

In the case of the C-metric, it exhibits a topological number $W=0$, which transitions to $-1$ in AdS space \cite{Wu:2023meo}. When incorporating black hole charge or spin, such as RN-, Kerr-, KN-C-metrics, the topological number becomes $W=1$, whereas it remains $W=0$ in the corresponding AdS space \cite{Wu:2023meo, Liu:2025iyl}.

The Taub-NUT solution differs from the C-metric. In its basic form, without additional black hole parameters, it possesses a topological number of $W=-1$. However, upon introducing black hole charge, angular momentum, and cosmological constant, the topological number transforms to $W=0$ \cite{Wu:2023xpq, Wu:2023fcw}. A positive topological number is only observed in RN-NUT-AdS black holes that satisfy the condition $Q^2 \geq n^2 + 3n^4/l^2$ (where $n$ represents the NUT charge) \cite{Azreg-Ainou:2025hpa}, or when the black hole carries a substantial magnetic charge \cite{Shahzad:2024yrx}.

\subsection{Cosmological constant}

Motivated by the AdS/CFT correspondence, black holes with negative cosmological constants have garnered significant attention. In various gravity theories such as Gauss-Bonnet gravity, massive gravity, quasitopological gravity, $f(R, T)$ gravity, rainbow gravity, and others, the coupling parameters and black hole characteristics can influence the number and positions of zero points in vector spaces. However, the total topological number typically remains constant across most scenarios \cite{Liu:2022aqt, Bai:2022vmx, Li:2023ppc, Li:2023men, Ali:2023jox, Du:2023heb, Shahzad:2023cis, Shahzad:2024ycq, Chen:2024kmy, Chen:2024atr, Hazarika:2024dex, Wang:2024fhf, Jeon:2024yey, Ma:2024ysf, Chen:2024ilt, Sadeghi:2024moo, Shahzad:2024pti, NooriGashti:2024bbc, TranNHung:2025omc, Yin:2025rhr}.

Conversely, when a positive cosmological constant is present, a cosmological horizon emerges. This leads to challenges in defining the corresponding thermodynamics due to the distinct black hole thermodynamics existing on the black hole horizon and cosmological horizon. Unless under the Nariai limit, where the temperatures on both horizons coincide, the system appears to reach thermodynamic equilibrium. While the small-large black hole transition differs from that in AdS space \cite{ZhangZhao}, it provides an avenue to explore thermodynamic topology.

In the context of thermodynamics residing on the black hole event horizon, the topology remains unaffected. For instance, the Schwarzschild-dS black hole possesses a topological number of $W=-1$, while RN-dS, Kerr-dS, and KN-dS black holes exhibit a vanishing topological number. On the other hand, for thermodynamics situated on the cosmological horizon, the topological number switches to $W=1$, while other cases witness no change \cite{Du:2023wwg}.

\subsection{Ensembles}

As we are aware, the charged AdS black hole only undergoes a small-large black hole phase transition in the canonical ensemble, where the black hole charge remains fixed. However, if the electric potential is held constant, no small-large black hole phase transition occurs; instead, the system experiences the Hawking-Page phase transition within the grand canonical ensemble. This raises the question of whether the behavior of the topological number varies across different ensembles.

In the case of the Dyonic AdS black hole, characterized by both electric and magnetic charges, three distinct ensembles are considered. In the canonical and grand canonical ensembles, these charges or their respective potentials are fixed. The mixed ensemble scenario involves a constant magnetic charge and electric potential. Notably, a study in Ref. \cite{Gogoi:2023xzy} revealed that in both the canonical and mixed ensembles, a conventional critical point exists with a topological number of -1, whereas no critical point is identified in the grand canonical ensemble. This discrepancy indicates that different ensembles present varying topologies for critical points.

Regarding the topology of the black hole solution, the topological number is $W=1$ for both the canonical and mixed ensembles. However, in the grand ensemble, the topological number is $W=0$ when $\phi_{e}^{2} + \phi_{m}^{2} < 1$ and $W=1$ when $\phi_{e}^{2} + \phi_{m}^{2} > 1$, where $\phi_{e}$ and $\phi_{m}$ represent the electric and magnetic potentials, respectively.

In the study of the Euler Heisenberg (EH)-AdS black hole, results are presented in Ref. \cite{Gogoi:2023wih}. In the canonical ensemble, the topological number is $W=1$ for negative EH parameters and $W=0$ for positive parameters. Regarding the higher-order QED corrected EH-AdS black hole, the topological number consistently remains $W=1$, irrespective of the EH parameter $a$. In the grand canonical ensemble, the values of the EH parameter do not impact the topological number, which maintains a value of $W=0$.

Comprehensive results are also available in Refs. \cite{Chen:2023pqk, Hazarika:2023iwp, Liu:2023sbf, Baruah:2024ovw}. Across these results, it is observed that the same black hole solution can be categorized into different classes based on the specific ensemble considered.

\subsection{Regular black holes}

In general relativity, black holes harbor a singularity at their core. To circumvent this issue, a class of regular black holes has been proposed, ensuring a well-behaved spacetime at $r=0$. Two primary methods are employed to construct these regular black holes: one involving a nonlinear electromagnetic field and the other stemming solely from pure gravity.

Early models such as the Bardeen and Hayward black holes \cite{Bardeen, Hayward} have elucidated this concept, wherein the inclusion of a nonlinear electromagnetic field term in the action aids in their interpretation \cite{Ayon-Beato, FanWang}.

Exploring regular Bardeen black holes within the context of negative cosmological constants, quintessence, massive terms, and Gauss-Bonnet terms, study indicates that the topological number remains $W=1$ when considering the black hole as a defect in thermodynamic parameter space \cite{Sadeghi:2023aii}. Conversely, for critical point topologies, Hayward black holes consistently exhibit $W=-1$, irrespective of the negative cosmological constant, quintessence, or Gauss-Bonnet terms \cite{Sadeghi:2024krq}. Notably, the topological number for Hayward black holes surrounded by string fluids is notably influenced by fluid parameters \cite{Anand:2025mlc}. Further insights into other regular black holes or black strings can be found in references such as \cite{Hung:2023ggz, Shahzad:2024ojx, TranNHung:2024tyg, Zhang:2024zqd}.

An alternative approach to constructing regular black holes involves pure gravity, where an infinite series of higher-curvature corrections is incorporated into the action \cite{BuenoCano}. The Bardeen and Hayward black holes can also be derived using this methodology; however, its applicability is limited to dimensions $d\geq 5$. In a recent study \cite{Wang:2024zlq}, the question arose as to whether these regular black holes, fashioned from pure gravity, could be grouped within the same thermodynamic topology.

Through the examination of two specific types of regular black holes, it was determined that the topological number reduces to zero, i.e., $W=0$. Moreover, upon further analysis of the general construction process, it was observed that all these regular black holes share the same topological number. This discovery unveils a universal topological characteristic exhibited by regular black holes arising from pure gravity.

\subsection{Residue approach}

The topological number is also related to the residue theorem, employed with which the topological properties of different black hole can be calculated \cite{Fang:2022rsb}.

The key point is defining a complex function
\begin{equation}
\mathcal{R}(z)\equiv\frac{1}{\tau-\mathcal{G}(z)},
\end{equation}
where $\mathcal{G}(r_h)=\tau$. The winding number $w_i$ of a singular point $z_i$ is given by
\begin{equation}\label{wind}
w_i =\frac{\textrm{Res}\mathcal{R}(z_i)}{|\textrm{Res}\mathcal{R}(z_i)|}=\textrm{Sgn}[\textrm{Res}\mathcal{R}(z_i)],
\end{equation}
where $|\,|$ denotes the absolute value of the complex function, and $\textrm{Sgn}(x)$ represents the sign function---this function returns the sign of a real number and yields 0 when $x=0$. It should be noted that since the singular points of interest are real, $\textrm{Res}\mathcal{R}(z_i)\in\mathbb{R}$. The global topological number $W$ of the black hole spacetime can thus be obtained as
\begin{equation}
W=\sum_i w_i.
\end{equation}
For the Schwarzschild black hole, it admits only one singular point, and gives $W=-1$. For the RN and RN-AdS black holes, the topological number is $W=$ 0 and 1, respectively. This result is the same as that given in Ref. \cite{WeiLiuad2022}.

\subsection{Statistical mechanics}

In general relativity, the entropy of a black hole amounts to a quarter of its event horizon area, a unique characteristic of gravity that sets it apart from conventional thermodynamic systems. Nevertheless, certain aspects of statistical mechanics have garnered significant interest, particularly in scenarios where the extensive entropy undergoes corrections, consequently impacting the system's topology.

Below, we present some examples of non-extensive entropies:

\textbf{R\'{e}nyi Entropy} \cite{Renyi}:
\begin{eqnarray}
\tilde{S}_R = \frac{1}{\lambda} \ln(1 + \lambda S_{BH}).
\end{eqnarray}
Here, the R\'{e}nyi parameter $\lambda$ is restricted to the range (0, 1) for a black hole system. As $\lambda$ approaches 0, the classical Boltzmann-Gibbs statistics is recovered.

\textbf{Barrow Entropy }\cite{Barrow}:
\begin{eqnarray}
\tilde{S}_B = \left(\frac{A}{4l_{P}^2}\right)^{1 + \Delta}.
\end{eqnarray}
It is important to note that the exponent $\Delta$ may vary in definition across different studies. Upon setting $\Delta$ to 0, the Bekenstein-Hawking entropy emerges.

\textbf{Sharma-Mittal Entropy} \cite{Masi}:
\begin{eqnarray}
\tilde{S}_{SM} = \frac{1}{\alpha} \left((1 + \beta S_{T})^{\frac{\alpha}{\beta}} - 1\right).
\end{eqnarray}
Here, $S_{T}$ represents Tsallis entropy. The parameters $\alpha$ and $\beta$ are determined by experimental data and serve as two phenomenological factors.

By substituting the entropy in (\ref{grf}) with these non-extensive entropies $\tilde{S}$, the respective topology can be determined through the generalized free energy:
\begin{eqnarray}
\mathcal{F} = M - \frac{\tilde{S}}{\tau}.
\end{eqnarray}
The majority of studies addressing this matter indicate a significant alteration of the zero points through such modifications. However, the topological number remains invariant, indicating that these systems continue to fall within the same topological classifications \cite{BahrozBrzo:2025pgt, Yasir:2024wir, Sekhmani:2024lsd, Zafar:2025sxl, Aslam:2025ren, Mohamed:2024kyb, Hazarika:2024xki, Tong:2023kob, Zhang:2023svu, Barzi:2023msl, Barzi:2024tqf, Yasir:2025uov, NooriGashti:2024dvq, Mohamed:2024plq, NooriGashti:2024tog, NooriGashti:2024ywc, NooriGashti:2025xfg, Hazarika:2024lnx, Ladghami:2025rkc}. However, certain exceptional cases have been identified. For instance, as demonstrated in Ref. \cite{Afshar:2025sav}, the topological number for the Einstein-Gauss-Bonnet theory is contingent on the specific statistical mechanics being considered. In the context of the Sharma-Mittal entropy, the topological number is $W=0$ for $\alpha \leq \beta$, while $W=1$ for $\alpha > \beta$. This indicates that the topological number may exhibit variability across different statistical mechanics approaches.

\subsection{Multiple defect curves}

For a given black hole solution, the defect curve forms a continuous function with respect to $\tau$ when the other black hole parameters are held constant. Notably, the parameter $\tau$ exerts no influence on the topological number. The emergence of multiple distinct defect curves can present an intriguing scenario, as explored in the context of dyonic black holes where both electric and magnetic charges are considered \cite{LiuLu}.

These black holes manifest a triple point and separated coexistence curve \cite{LiWangWei}. To delineate the existence of multiple defect curves, the corresponding parameter space is examined. Choosing the coupling parameters $\alpha_1=0.41$ and $\alpha_2=50$, three illustrative instances with $Pr_{0}^2$ values of 0.05, 0.0003, and 0.00004 are investigated \cite{Chen:2024sow}. In the initial case, a single defect curve is observed, yielding a topological number of $W=1$. As the pressure decreases, exemplified by $Pr_{0}^2=0.0003$, a new defect curve emerges near the small black hole horizon. For each fixed $\tau/r_0$, up to three zero points are identified for the constructed vector. Notably, the novel additions exhibit opposing winding numbers, yet the topological number remains constant. Subsequently, with a further reduction in pressure, specifically in the third scenario with $Pr_{0}^2=0.00004$, five zero points are discerned for certain $\tau/r_0$ values. Despite this increase, the topological number still retains its value at $W=1$.

Consequently, it is evident that the topological number remains unaffected by the presence of multiple defect curves.

\subsection{$\tau$ dependent: an exception}

Previous study has consistently shown that the topological number remains constant with respect to $\tau$ when the other parameters of a black hole are fixed. Yet, an intriguing departure from this norm is highlighted in a recent study \cite{Wu:2024rmv}, featuring static multi-charge AdS black holes within gauged supergravity. In this scenario, if two electric charge parameters hold non-zero values, the thermodynamic topological number will exhibit a notable temperature-dependent characteristic. For example, as $\tau$ increases, the topological number may undergo a transition from 1 to 0. This phenomenon primarily stems from the temperature converging towards a finite value as the black hole's event horizon contracts to zero, a distinctive attribute unique to these black holes.

This topological analysis spans various black hole solutions, encompassing a diverse range of systems and scenarios. Given the unique attributes of each system, a detailed examination of individual cases will not be provided here. For in-depth information on specific systems, we direct readers to the corresponding references \cite{Bai:2022klw,Chatzifotis:2023ioc,Du:2023nkr,Fairoos:2023jvw,Gogoi:2023qku,Zhang:2023tlq,
Chen:2023ddv,Sadeghi:2023dxy,Chen:2023elp,Malik:2023nso,Rizwan:2023ivp,Fairoos:2023hkk,
Saavedra:2023lds,Sadeghi:2023dsg,Hazarika:2024cpg,Wang:2024zbp,Wu:2024txe,Lei:2024wvj,Li:2024ial,EslamPanah:2024fls,
Hazarika:2024xar,Sekhmani:2024vsu,He:2024jho,Hazarika:2024cji,Rathi:2024ycw,Shi:2024wwz,
Dong:2024hod,Sajadi:2025prp,Pantig:2025paj,Anand:2025qow,AraujoFilho:2025hnf,Dong:2025qnp,Anand:2025vfj,Nieto:2025apz,Zhang:2025cev,Fan:2022bsq,Zhang:2023uay,Alipour:2023uzo,Xu:2023vyj,
Sadeghi:2023tuj,Wang:2023qxw,AnandGashti,Panotopoulos,BaoMa,AhmedGashti,MaHuo,SekhmaniLuciano,EsmailiEsmaili,AiAi,AhmNasheded,AhBaomed,HazarikaPhukon,RaniRiaz,SekhmaniMaurya,
SekhmaniMauryaRayimbaev,GogoiGogoi,GashtiAfshar,HosseinifarHosseinifar,AlipourAlipour,SekhmaniSekhmani,WuWu,NamNam,ChenChenChen}.

\section{Universal topological classifications}
\label{Universal}

As demonstrated earlier, the black hole parameters and the variable $\tau$ play a significant role in determining the number and local winding characteristics of zero points within the constructed vector when considering black hole solutions as defects. However, with only a few exceptions, they do not impact the global topological number. This observation implies the potential development of a theory encompassing universal topological classifications. Such a theory could categorize diverse black holes into finite classes, enabling the examination of universal properties within these classifications.

\subsection{Topological classifications}

To determine the global topological number, a contour encompassing the entire parameter space defined by $\Theta$ and $r_h$ must be constructed. The parameter $\Theta$ is constrained within the range of (0, $\pi$), while the horizon radius $r_{h}$ falls within the interval $(r_{m},~\infty)$, where $r_m$ represents the minimum value of $r_{h}$. For instance, in the case of the Schwarzschild black hole, $r_m=0$, whereas for the RN black hole, $r_m=M=Q$, corresponding to the extremal black hole.

At $r=r_m$ and as $r$ tends to infinity, we assume that the inverse of the temperature exhibits the following asymptotic behaviors
\begin{eqnarray}
 &case\; I\;:& \beta(r_{m})=0,\quad\;\; \beta(\infty)=\infty,\label{a1}\\
 &case\; II:& \beta(r_{m})=\infty,\quad \beta(\infty)=\infty,\\
 &case\; III:& \beta(r_{m})=\infty,\quad \beta(\infty)=0,\\
  &case\; IV:& \beta(r_{m})=0,\quad\;\; \beta(\infty)=0. \label{a4}
\end{eqnarray}
The contour encompassing all possible parameter regions is typically defined as $C=I_1\cup I_2\cup I_3\cup I_4$, where $I_{1}=\{r_{h}=\infty,\;\Theta\in(0, \pi)\}$, $I_{2}=\{r_{h}=(\infty,\;r_{m}),\;\Theta=\pi\}$, $I_{3}=\{r_{h}=r_{m},\;\Theta\in(\pi, 0)\}$, and $I_{4}=\{r_{h}=(r_{m},\;\infty),\;\Theta=0\}$. To calculate the topological number for the contour $C$, it is necessary to analyze the behavior of $\phi$ along these segments. The results are summarized in Table I of Ref. \cite{Wei:2024gfz}.

\begin{table}[]
\setlength{\tabcolsep}{2.5mm}{\begin{tabular}{cccccc}\hline\hline
     & $I_{1}$ &$I_{2}$ & $I_{3}$ & $I_{4}$ &$W$ \\\hline\hline
 case I      & $\leftarrow$ & $\uparrow$& $\rightarrow$ & $\downarrow$&-1 \\
 case II   & $\leftarrow$   &$\uparrow$ & $\leftarrow$     &$\downarrow$ &0 \\
 case IIII      &  $\rightarrow$   & $\uparrow$& $\leftarrow$ & $\downarrow$&+1\\
 case IV   & $\rightarrow$ &$\uparrow$ & $\rightarrow$ & $\downarrow$&0 \\\hline\hline
\end{tabular}
\caption{The directions indicated by the arrows of $\phi^{r_{h}}$ are shown for the four segments. The corresponding topological number for each case is also listed in the last column \cite{Wei:2024gfz}.}\label{tab}}
\end{table}

Evidently, the topological number $W$ takes values of 0, $\pm$1, suggesting three distinct topological classifications. However, as proposed in Ref. \cite{Wei:2024gfz}, although case II and case IV share the same topological number, they exhibit contrasting thermodynamic behaviors. Notably, the RN black hole and the Schwarzschild-AdS black hole serve as quintessential examples of case II and IV, respectively. The RN black hole features a maximum temperature, whereas the Schwarzschild-AdS black hole showcases a minimum temperature. Furthermore, the RN black hole includes an extremal configuration with the horizon at its minimum radius, while the horizon radius of the Schwarzschild-AdS black hole can diminish to zero. Despite both having two zero points in the parameter space vector, the sequence of their occurrence differs. Given these disparate thermodynamic characteristics, cases II and case IV should be categorized into distinct classes. Consequently, four topological classifications for black hole thermodynamics are introduced as
\begin{eqnarray}
W^{1-}, \quad W^{0+}, \quad W^{0-}, \quad W^{1+}, \label{tttp}
\end{eqnarray}
corresponding to topological numbers -1, 0, 0, 1, respectively. The indices $0+$ and $0-$ signify that the first zero point of the vector possesses a positive or negative winding number.

The RN and Schwarzschild-AdS black holes serve as straightforward examples. However, in the case of certain other black hole solutions, the constructed vector may exhibit more than two zero points. Consequently, as the horizon radius increases, a sequence of zero points with varying winding numbers emerges. Typically, positive and negative winding numbers manifest alternately in accordance with the topological construction. For the four classes outlined in Eq. (\ref{tttp}), as the horizon radius increases, the systematic arrangement of the winding numbers corresponding to these zero points follows the patterns of [-, (+, -), ..., (+, -)], [+, (-, +), ..., -], [-, (+, -), ..., +], and [+, (-, +), ..., (-, +)], respectively. To differentiate these cases, one can utilize the signs of the innermost and outermost winding numbers:
\begin{eqnarray}
 W^{1-}=[-, -], \quad W^{0+}=[+, -], \nonumber\\
 W^{0-}=[-, +],  \quad  W^{1+}=[+, +].\label{aidos}
\end{eqnarray}
Consequently, a notation is proposed for the topological classifications, enabling the categorization of each black hole solution into these classes.

\subsection{Universal thermodynamical behaviours}

Once these systems have been classified, delving into their universal thermodynamic behaviors becomes essential. Four important limits necessitate examination. Two of these limits pertain to the innermost and outermost states, representing the scenarios of small and large black holes, respectively. The remaining two limits correspond to low and high temperature regimes.

In the classifications of $W^{1-}$, $W^{0+}$, $W^{0-}$, and $W^{1+}$, the innermost black hole states are delineated as unstable, stable, unstable, and stable, while the outermost black hole states are characterized as unstable, unstable, stable, and stable, respectively.

In the low-temperature limit as $\beta\to\infty$, the $W^{1-}$ class presents a large black hole that demonstrates thermodynamic instability due to its topological number. In contrast, the $W^{1+}$ class features a single stable small black hole state. Within the $W^{0+}$ class, an unstable large black hole state coexists with a stable small black hole state. Notably, the $W^{0-}$ class does not exhibit any black hole state at the low temperature limit.

As the high-temperature limit approaches with $\beta\to 0$, the $W^{0+}$ class is distinctive for the absence of any black hole state, while the $W^{0-}$ class demonstrates both an unstable small black hole state and a stable large black hole state. In the $W^{1-}$ class, one encounters an unstable small black hole state, whereas the $W^{1+}$ class is characterized by a stable large black hole state.

In conclusion, the universal thermodynamic behaviors have been investigated based on the topological classifications. For clarity, the summarized results can also be referenced in Table II of the study given in Ref. \cite{Wei:2024gfz}.

Furthermore, the universal classifications for various black hole solutions have been explored in other works \cite{Zhu:2024zcl, Rizwan:2025pvp, Chen:2025nto, Chen:2025fse}. Note that an intriguing example is the observation of a topological phase transition between the $W^{0-}$ and $W^{1+}$ classes \cite{Wu:2024asq}. In addition, certain other black hole systems with richer phase structures, such as those exhibiting superfluid phase transitions \cite{Tjoa2} and multi-critical points \cite{Tavakoli}, deserve further exploration.

\section{Discussions and conclusions}
\label{Conclusion}

In this work, we briefly reviewed the development of thermodynamic topologies concerning critical points, Davies points, the Hawking-Page phase transition, and black hole solutions. By constructing vectors wherein the zero points signify defects within the thermodynamic parameter space, including critical and Davies points, we ascertained the winding number and topological number. Consequently, a comprehensive thermodynamic topology, complete with specific notation, was delineated.

The treatment of black hole solutions as topological defects has garnered significant attention. This interest is partly due to the fact that critical points, Davies points, and the Hawking-Page phase transition only manifest in select black hole solutions. Consequently, viewing black hole solutions as defects holds universal interest, in which local thermodynamic stability is considered a topological property. The presence of a positive or negative winding number signifies the stability or instability of a black hole. Moreover, an intriguing aspect emerges as topological classifications were formulated, categorizing different black hole solutions accordingly. Notably, the universal thermodynamic properties were explored concerning small and large black holes, as well as the low and high temperature limits.

The utilization of topological methods has emerged as a robust testing ground within black hole physics. Through the topological approach, phenomena such as the photon sphere/light ring, the deflection of light, time-like circular orbits, and the Hawking temperature can be effectively examined \cite{Cunhab,Wei:2022mzv,GibbonsWerner,RobsonVillari,Wei:2023bgp,Ye:2024sus}. Exploring the interplay of topologies in both gravity and thermodynamics represents a significant pursuit. This exploration presents a unique opportunity to unveil the nature of black holes and spacetime from a thermodynamic perspective.

\section*{Acknowledgements}
This work was supported by the National Natural Science Foundation of China (Grants No. 12475055, No. 12475056, and No. 12047501), the Fundamental Research Funds for the Central Universities (Grant No. lzujbky-2025-jdzx07), the Natural Science Foundation of Gansu Province (No. 22JR5RA389, No.25JRRA799).

~\\

\textbf{Conflict of Interest}  The authors declare that they have no conflict of interest.

\end{document}